\documentclass[12pt,preprint]{aastex}

\newcommand{\kms}{km~s$^{-1}$}
\newcommand{\Hone}{H\,{\sc i}}
\newcommand{\Htwo}{H\,{\sc ii}}
\newcommand{\Ha}{H$\alpha$}

\newcommand{\Stwo}{[S\,{\sc ii}]}
\newcommand{\Msun}{M$_{\sun}$}
\newcommand{\Lsun}{L$_{\sun}$}

\newcommand{\vlsr}{v$_{LSR}$}

\begin{document}

\title{A VLA Search for Water Masers in Six \Htwo\ Regions:
Tracers of Triggered Low-Mass Star Formation}

\author{Kevin R. Healy and J. Jeff Hester}
\affil{Department of Physics and Astronomy}
\affil{Arizona State University, Tempe, Arizona 85287}
\email{kevin.healy@asu.edu, jeff.hester@asu.edu}

\and

\author{Mark J. Claussen}
\affil{National Radio Astronomy Observatory Array Operations Center}
\affil{P.O. Box 0, 1003 Lopezville Road, Socorro, NM 87801}
\email{mclausse@aoc.nrao.edu}

\begin{abstract}

We present a search for water maser emission at 22 GHz associated with
young low-mass protostars in six \Htwo\ regions --- M16, M20,
NGC~2264, NGC~6357, S125, and S140.  The survey was conducted with the
NRAO Very Large Array from 2000 to 2002. For several of these \Htwo\
regions, ours are the first high-resolution observations of water
masers. We detected 16 water masers: eight in M16, four in M20, three
in S140, and one in NGC~2264.  All but one of these were previously
undetected.  No maser emission was detected from NGC~6357 or S125.
There are two principle results to our study.  (1) The distribution of
water masers in M16 and M20 does not appear to be random but instead
is concentrated in a layer of compressed gas within a few tenths of a
parsec of the ionization front.  (2) Significantly fewer masers are
seen in the observed fields than expected based on other indications
of ongoing star formation, indicating that the maser-exciting lifetime
of protostars is much shorter in \Htwo\ regions than in regions of
isolated star formation.  Both of these results confirm predictions of
a scenario in which star formation is first triggered by shocks driven
in advance of ionization fronts, and then truncated $\sim 10^5$ years
later when the region is overrun by the ionization front.

\end{abstract}

\keywords{masers --- ISM: individual (M16, M20, NGC 2264, NGC 6357, S125, S140) --- stars: pre-main sequence --- stars: formation}

\section{Introduction}

The majority of low-mass stars form not in regions of isolated star
formation, such as the Taurus-Auriga molecular cloud, but instead in
cluster environments that also contain massive stars \citep[see,
e.g.][]{ll03}. The process of star formation may be fundamentally
different in these two environments. In low-mass star-forming
environments, each protostar is largely responsible for shaping its
own natal environment.  In contrast, star-forming environments
surrounding \Htwo\ regions are shaped by the radiation and mechanical
energy of the O and B stars that excite the region.  Much effort has
been made to understand the importance of supernova shocks as triggers
of low-mass star formation \citep[see, e.g.][]{fb96}.  But an even more
important effect may involve triggering of star formation by
radiatively driven implosion of molecular gas surrounding \Htwo\
regions \citep[see, e.g.][]{el77,ber89,bm90}.  \citet{hes96} pointed
out that such regions would be quickly overrun by the advance of the
ionization front, exposing cloud cores and protostars to the
potentially disruptive environment in the interior of the \Htwo\
region.  Recent optical and near-infrared observations of young
stellar objects at the edges of \Htwo\ regions
\citep[e.g.][]{ma02,osp02,sug02,tsh02} provide additional support for
the triggering and truncation of the star formation process in these
environments.

Although the stellar population in the low density interior of the
\Htwo\ region can be studied in the optical and near-infrared part of
the spectrum, the youngest protostars are extremely difficult to
observe at these wavelengths.  This is due in part to the high
foreground extinction associated with the molecular gas in which the
protostars are embedded.  This is also due to the fact that protostars
are intrinsically cool objects, radiating at centimeter wavelengths
and peaking at submillimeter wavelengths \citep{awb93}.  If we are to
address properly the question of low-mass star formation in complex
\Htwo\ region environments, then we need a way to probe the spatial
distribution of young protostars, including low-mass protostars, in
the high-density molecular gas bounding \Htwo\ regions.  In this paper
we present observations of water masers in and around \Htwo\ regions,
and find such observations to provide a powerful yet previously
under-utilized tool with which to conduct such studies.

Previous observations of water maser emission from low-mass protostars
have focused on the relatively nearby sample selected from low-mass
star-forming regions like Taurus-Auriga.  For example, the sample of
\citet{cla96} was selected to have d $\lesssim$ 450 pc and 90\% of the
sources of \citet{fur01} had d $\lesssim$ 500 pc.  These and other
studies have provided a wealth of data on the behavior of water maser
emission.  We would like to apply that knowledge to the more distant
and more complex case of star formation and water maser activity in
regions shaped by the presence of massive stars.

Water maser emission is only known to arise in association with AGN,
AGB stars, and protostars.  For example, \citet{cla99} find that even
the most favorable conditions behind supernova remnant shocks fail to
produce water maser activity.  Maser emission in the
$6_{1,6}\rightarrow5_{2,3}$ transition of water at 22.235077 GHz is a
well-known trait of the youngest protostars, low- and high-mass alike
\citep{tof95,cla96,mee98,fur03}.  Observational studies of both
massive and low-mass protostars suggest that maser activity is
associated with the youngest stages of star formation.
\citet{fur01,fur03} show that $\sim$40$\pm$20\% of Class 0 and
$\sim$4$\pm$3\% of Class I low-mass protostars exhibit water maser
emission.  \citet{fc00} conclude that water masers are likely excited
by massive pre-main sequence stars or possibly actively-accreting,
massive protostars.  Observations also show that water maser activity
lies in close proximity to the exciting object.  Water masers are
observed within tens or hundreds of AU of low-mass protostars
\citep[e.g.][]{che95,cla98,fur00}. The separation between maser and
massive star can be $\lesssim10^4$ AU \citep[e.g.][]{tof95,fc00}.
Only in extreme cases (e.g. the water masers in Orion), does the
spatial correlation between water masers and the exciting protostar
break down. In general, the presence of water masers in a region of
star formation, such as the molecular gas surrounding an \Htwo\
region, is an unambiguous indication of the presence of a star in the
earliest stages of formation. Due to the intrinsic variability and
emission geometry of maser emission, this detection method will not
uncover all young protostars.  Nonetheless, water maser emission can
provide us with a look at the distribution of young protostars within
an \Htwo\ region environment. Finally, while low- and high-mass
protostars may result from the star formation around \Htwo\ regions,
far more low-mass stars form in this process \citep{ll03}.
Statistically we expect that most of the observed water maser activity
in \Htwo\ regions is excited by low-mass protostars.

The presence of water masers in \Htwo\ region environments is
well-known \citep[see, e.g.][]{gd77,bl79,com90}. These surveys and
more recent searches have been performed with single-dish telescopes,
which have typically 1 arcminute resolution.  This resolution is
inadequate to properly locate protostars relative to ionization fronts
and other details of the \Htwo\ region environment.  If we want to use
water maser emission to study the process of star formation in regions
around massive stars, we need positions accurate to 1\arcsec\ or better,
comparable to optical and near-infrared observations of \Htwo\
regions.  The NRAO Very Large Array (VLA)\footnote{The National Radio
Astronomy Observatory is a facility of the National Science Foundation
operated under cooperative agreement by Associated Universities, Inc.}
is the instrument of choice for this type of investigation compared to
single-dish telescopes. First, the greater effective aperture of the
VLA provides greater sensitivity which allows for weaker masers to be
detected in shorter observations compared to single-dish
observations. Second, the large primary beam of the VLA allows for
fewer pointings to cover the same area of sky compared to large
aperture single-dish telescopes. Third, the small synthesized beam of
the VLA gives a precise position for water masers that can be compared
to optical or infrared observations of the \Htwo\ regions.

The observations, calibration, and analysis of this survey are
discussed in $\S2$.  The results of the observations for each \Htwo\
region are presented in $\S3$. In $\S4$, we discuss our results, and
find that we are able to test two specific predictions concerning
triggering and subsequent disruption of low-mass star formation in
these environments.  A summary of our results and conclusions is
presented in $\S5$.

\section{Observations and Data Reduction}

We searched the \Htwo\ regions M16, M20, NGC~2264, NGC~6357, S125, and
S140 for water maser emission at 22.2 GHz. Our original observations
of four \Htwo\ regions (NGC~2264, M16, M20, and S140) were made on
2000 April 02, April 04, June 06, and June 13 (C configuration,
angular resolution 0$\farcs$9). We expanded our observations of M16
and M20 and added NGC~6357 and S125 on 2001 March 26 and May 18 (B
configuration, angular resolution 0$\farcs$3). A final set of
observations of select targets within M16 and M20 occurred on 2002
March 14 (A configuration, angular resolution 0$\farcs$08). Due to the
intrinsic variability of maser emission, we planned to observe each
location within the target \Htwo\ regions at least twice, separated by
a few months' time. This was achieved for all of the targets, except
S140, which was observed only on 2000 June 06.

A total bandwidth of 3.125 MHz was divided into 127 channels, giving
24.4 kHz (0.33 \kms) channel width. For the 2000 and 2001
observations, we used one frequency band centered on the systemic
velocity of the \Htwo\ regions. For the 2002 observations, we used two
frequency bands (total bandwidth 6.25 MHz) that overlapped by several
channels. In this mode, the combined bandwidth of the overlapping
frequency bands was centered on the systemic velocity of the \Htwo\
regions. The pointings, dates of observation, and velocity range
searched for each \Htwo\ region are listed in Table~\ref{observed}.

\begin{figure}
\plotone{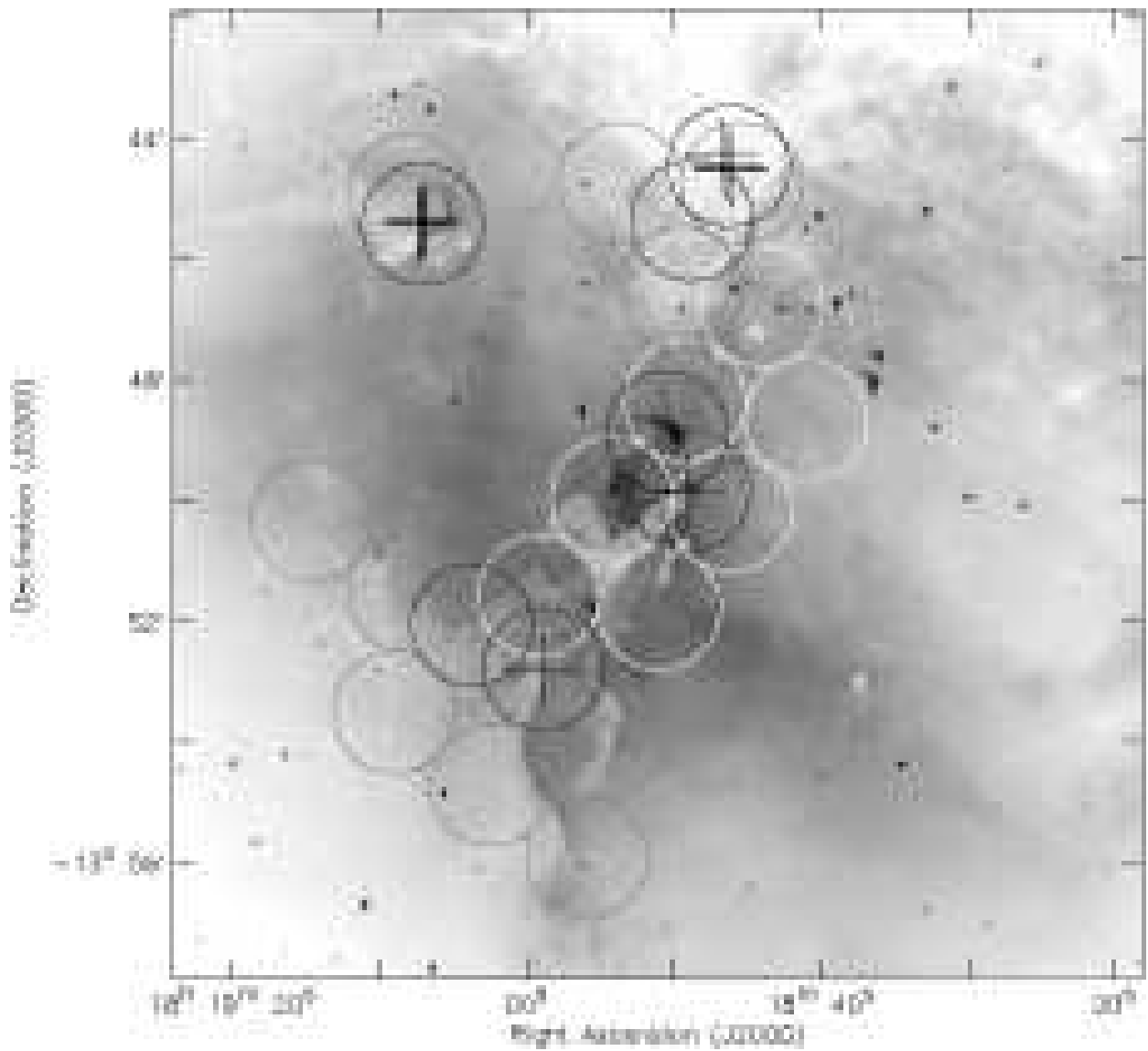}
\caption{\label{m16finder} Pointing map and maser locations for M16.
Palomar 1.5-meter \Ha\ image (greyscale) overlaid with positions of
VLA pointings (circles: white - 2000; grey - 2001; black - 2002) and
maser detections (crosses). The circle diameters are equal to the VLA
primary beam FWHM at the observing frequency of 22.2 GHz.}
\end{figure}

At 22 GHz, the primary beam of VLA antennae is 2$\arcmin$ FWHM.
We used a patchwork of pointings to search portions of the \Htwo\
regions for water maser emission. In our 2000 observations, our
pointings were contiguous and covered a small portion of the four
target \Htwo\ regions.  In 2001 and 2002, we observed a variety of
targets of interest as well as covering a larger portion of the \Htwo\
regions (see Figures~\ref{m16finder}-\ref{n6357finder}).  All
pointings, but one, was observed for 6 to 7 minutes. The exception,
NGC~2264 pointing 1 (2000 April 04), was observed for just over one
minute.

We calibrated the data in the usual way with the Astronomical Image
Processing System (AIPS), using 0137+331 (3C48) as the primary flux
calibrator for all observations, except 2002 March 14 for which we
used 1331+305 (3C286). The phase calibrators were 0700+171 (NGC~2264),
1733-130 (M16 and NGC~6357), 1820-254 (M20), 1911-201 (M20), 2022+616
(S140), and 2202+422 (S125). The derived fluxes are listed in
Table~\ref{calib} and are in agreement with those reported by other
observers as listed in the online VLA calibrator
manual\footnote{http://www.aoc.nrao.edu/$\sim$gtaylor/calib.html}.

For each pointing, the entire primary beam was imaged using the AIPS
routine IMAGR with a pixel size of 0$\farcs$2 for the 2000
observations, 0$\farcs$06 for the 2001 observations, and 0$\farcs$015
for the 2002 observations. In each case the pixel size was chosen to
be approximately one-fifth the synthesized beamwidth. The resulting
spectral-line cubes were searched for maser components by identifying
peaks that were more than 7$\sigma$ above the noise level of each
channel image. Peaks within $\sim$2$\arcsec$ of the image edges were
rejected as spurious given the lower sensitivity at the edge of the
primary beam. In addition to this test, we also checked for multiple
features weaker than 7$\sigma$ that appeared at the same position in
different channels. This second test identified no components that
were not detected in the first test. For fields that included maser
emission stronger than 10 Jy, we used the CLEAN method
\citep{hog74,cla80} to remove the effects of the point spread function
of the bright maser emission from the data. These CLEAN'ed data cubes
were then searched again using both search tests. No additional maser
components were identified.

For those fields in which maser emission was identified, spectral-line
cubes were made and CLEANed. Primary beam attenuation was corrected
using the AIPS routine PBCOR.  The resulting cubes had rms noise
values of $\sim$25-45 mJy/beam for the 2000 observations (the range of
values mostly due to varying weather between the four dates of
observation in this year), $\sim$25-40 mJy/beam (with a few as high as
50 mJy/beam) for 2001, and 10-20 mJy/beam for 2002.

The position of the peak of emission in each channel image was
compared throughout the range of channels in which emission was
identified. No significant variation of position versus frequency was
found, and no resolved emission was detected.  These two results
simplified the process of spectrum extraction since a single position
for each maser component was valid for all channels. For each maser, a
bounding box was chosen that just enclosed the synthesized beam in the
channel of strongest emission, which for the reason above also
included the emission in the other channels. A spectrum of the maser
component was then extracted from the image cube using the AIPS
routine ISPEC.

\section{Results}

Four of the six \Htwo\ regions exhibited water maser emission for a
total of 16 spatial components: eight in M16, four in M20, one in
NGC~2264, and three in S140. We define a component as a spatially
unresolved spot of emission even if it contains multiple spectral
features. It should be noted, however, that at resolutions of tens to
hundreds of milliarcseconds, individual velocity features are often
spatially distinct \citep[see, e.g.][]{cla98}.  Table~\ref{detected}
lists our designation of each maser, the J2000 position, the date of
detection, the flux(es) in Janskys and the velocity in \kms\ of the
strong spectral component(s).  We detected no maser emission in
NGC~6357 and S125 to a detection limit of 0.1 Jy (3$\sigma$). Of our
16 detections, only the water maser associated with S140 IRS~1 (S140
component A) was found in the literature.

We detected 9 of the 16 water masers in more than one epoch. No
significant proper motion was detected in any of these nine components
when taking the dimensions of the synthesized beams into account. The
single exception was M16 Column 5 component A, but there was no
consistent direction of the displacement with time so we could not
calculate a meaningful average velocity vector. Individual velocity
components appeared and disappeared on timescales ranging from months
to years. In contrast, M16 Column 5 component A displayed a stable
spectrum during three epochs spanning a year.

\subsection{M16}

M16 (S49, ``Eagle Nebula'') is excited by the cluster NGC 6611 which
contains a single early O-type star and several middle O-type stars
\citep{bmn99}. Several studies show that the stellar population has a
spread of age as broad as a few million years, with pre- to post-main
sequence members present \citep{hil93,dew97}. This \Htwo\ region is
well known for its columns of neutral gas, dubbed ``elephant trunks''
\citep[see, e.g.][]{hes96}, that project into the ionized volume. The
presence of young, partially enshrouded protostars at the tips of
several elephant trunks illustrates the ongoing star formation in M16
\citep{tsh02,ma02}. The distance to M16 is 2.6 kpc according to
\citet{dew97}, in agreement with earlier studies.

Water maser emission associated with W37 in the center of M16 was
previously reported by \citet{yng75}. Water masers have also been
reported to the east and west of M16 by \citet{bl79}. During our three
years of observations, we observed both a portion of the cluster
NGC~6611 as well as several elephant trunks and the interior of the
\Htwo\ region. We focused our attention on the tips of Columns 1 and 2
\citep{hes96}, the column to the southeast (referred to as Column 4 in
\citealt{ma01}), and the column to the northeast (here referred to as
``Column 5''). In addition, the detection by \citet{yng75} led us to
observe the optical extinction feature north and east of NGC~6611
(here referred to as the ``North Bay''). The positions of the VLA
pointings are shown in Figure~\ref{m16finder} and listed in
Table~\ref{observed}. We detected 8 water masers in M16: single
components in Column 2 and 4, three components in Column 5, and three
components in the North Bay (see Table~\ref{detected}). We did not detect
water maser emission from Column 1, Column 3, nor the members of the
cluster NGC~6611 to a limit of 0.1 Jy (3$\sigma$).

Water maser emission was detected at the tip of Column 2 in five
epochs from 2000 April 04 to 2002 March 14. This emission consisted of
a velocity component at a \vlsr\ of about +22 \kms, with two
additional velocity components appearing in 2001 March 26 only (see
Figure~\ref{col2specs}).  Figure~\ref{col2wfnic} shows the position of
this maser relative to the {\it HST} WFPC2 \Ha, \Stwo, and NICMOS
F110W images. The maser emission lies about 2$\farcs$7 away (P.A. =
113$\degr$) from the Class I/II object M16 ES-2 \citep{tsh02}. At a
distance of 2.6 kpc, a separation of 2$\farcs$7 corresponds to 7000
AU, which is much larger than the typical separations between low-mass
protostars and the water masers they excite. We conclude the Column 2
maser is not associated with M16 ES-2. Thus the water maser activity
must be excited by a previously unseen protostar. At this position
within Column 2, there is no near-infrared source seen \citep[][and
see Figure~\ref{col2jhk}]{ma01}, making it likely that this
maser-exciting protostar is a Class 0 source.  Such a source might in
principle be detected at radio or submm wavelengths, although
observations such as those of \citet{whi99} lack the spatial
resolution necessary to identify low-luminosity sources against the
bright and structured nebular emission present at this location.  For
example, these observations did not show the presence of the protostar
M16 ES-2 identified by \citet{tsh02}.  \citet{pil98} also failed to
identify the presence of ES-2 in their ISO observations, despite the
fact that this 20\Lsun\ protostar is bright enough to be easily
detected by ISO.  Higher resolution and sensitivity radio, submm, and
FIR observations of this position will be necessary to identify the
protostar that excites this water maser emission.

Water maser emission was also detected from the tip of Column 4 in
five epochs from 2000 April 04 to 2002 March 14. Like the Column 2
maser, the spectrum of the Column 4 maser was relatively simple with
only one or two velocity components at similar \vlsr. All detected
maser emission was in the range \vlsr\ = +23.4 to +28.3 during the
three years of observation (see Figure~\ref{col4specs}). In the {\it
JHK} mosaic by \citet{ma01}, two features are visible in the vicinity
of the maser emission (see Figure~\ref{col4jhk}).  The northern
infrared feature has a ``core plus jet'' morphology.  This feature is
spatially coincident with the maser emission to within the alignment
uncertainty of the radio and infrared data.  The maser positions at
this location also appear to be slightly spread out along the inferred
position angle of the jet.  Together, these strengthen the
identification \citep{ma01} of the northern infrared source as the
protostar responsible for the HH 216 outflow.  While the outflow has
broken out of the interior of Column 4, the protostar itself is still
embedded within Column 4.  The presence of water maser emission and a
strong outflow suggests that this is a Class 0 or I source \citep{fur01}.

We have labeled the elephant trunk to the east of NGC~6611 as ``Column
5.'' In 2001 and 2002, we detected a component (A) composed of two
strong spectral peaks that maintained nearly constant LSR velocities
of approximately +16 and +21 \kms\ over the time span of the
observations.  In 2002 only, we detected components B and C which were
weaker than component A and separated from component A by
$\sim$3$\farcs$5 (P.A. = -22$\degr$) and $\sim$3$\farcs$9 (P.A. =
+68$\degr$), respectively. Figure~\ref{col5specs} shows the spectra of
these three masers. The physical separations between the three
components are nearly $10^4$ AU. This large separation implies this
water maser activity is excited by multiple protostars. Close
inspection of the near-infrared image of \citet{ma01} reveals a small
bipolar reflection nebula and several ``red'' (i.e. large, positive
J-K and H-K color) objects in the vicinity of the water masers (see
Figure~\ref{col5jhk}). Further evidence of star formation at this
location comes from the presence of the MSX point source
G017.0335+00.7479, less than 5$\arcsec$ south of component A
\citep{pri01}. As with the Column 2 and Column 4 protostars, the
protostars in Column 5 are on the verge of being overrun by the
advancing ionization front. As discussed below, the concentration
of young protostars so near to ionization fronts is unlikely to be
coincidental.

The optical extinction feature north and east of NGC~6611 is labeled
the ``North Bay.'' In 2001 and 2002, we detected several maser
components on the western side of the North Bay, within dense gas
visible in near- and mid-infrared observations \citep[][and the {\it
ISO} mosaic of M16\footnote{Available through the {\it ISO} archive at
http://www.iso.vilspa.esa.es/}]{pil98,ma01}. The brightest component
(A) exhibited two velocity features detected in both 2001 epochs: a
narrow peak at \vlsr\ = +24.7 \kms\ and a broad peak at \vlsr\ = +15.7
\kms.  Component B exhibited very different spectra in 2001 May 18 and
2002 March 14, with single peaks at +14.8 \kms\ and +6.3 \kms\ on the
two dates, respectively. Component C was the weakest, detected only on
2002 March 14, with three velocity components less than 0.15 Jy
each. Figure~\ref{nbayspecs} shows the spectra for these masers. The
separation between individual maser components in the North Bay is
even larger than the Column 5 masers, again suggesting multiple
protostars at this location. It is unclear when these protostars will
be overrun by the ionization front because the geometry is unfavorable
for locating the protostars relative to the ionization front.

Water maser emission at \vlsr\ of +10 and +25 \kms\ was reported by
\citet{yng75} at the position of W37, approximately 2$\arcmin$ north
of the tip of Column 1. We found no emission at the \citeauthor{yng75}
position to a detection limit of 0.1 Jy for all five epochs we
observed. Both the proximity of the Column 2 and North Bay masers to
W37 and the similarity of maser velocities detected in this study and
cited by \citet{yng75} suggests emission from the Column 2 or North
Bay was misidentified as emission from W37.

\subsection{M20}

M20 (S30, NGC~6514, ``Trifid Nebula'') is a young symmetric \Htwo\
region centered on and powered by the late O-type star HD 164492A.
The region shows evidence for strong foreground extinction in addition
to the characteristic triad of dust lanes. T Tauri stars within the
\Htwo\ region \citep{rho01} and dense cloud cores in the surrounding
neutral gas \citep{cer98,lc00} have been detected. Another obvious
sign of star formation is the HH399 jet launched by an as yet
unidentified protostar lying within the dense cloud TC2 located
southeast of HD 164492A \citep{cer98,hes99}. Two other massive cloud
cores, TC3 and TC4, lie at the southwestern periphery of M20
\citep{cer98}.  \citet{lc00} interpreted the small size of M20 as a
result of the young dynamical age (0.3-0.4 Myr) for this \Htwo\
region. The most commonly quoted distance to M20 is 1.68 kpc
\citep{lco85}, but larger values have been suggested \citep[see,
e.g.][and references therein]{kml99}.

\begin{figure}
\plotone{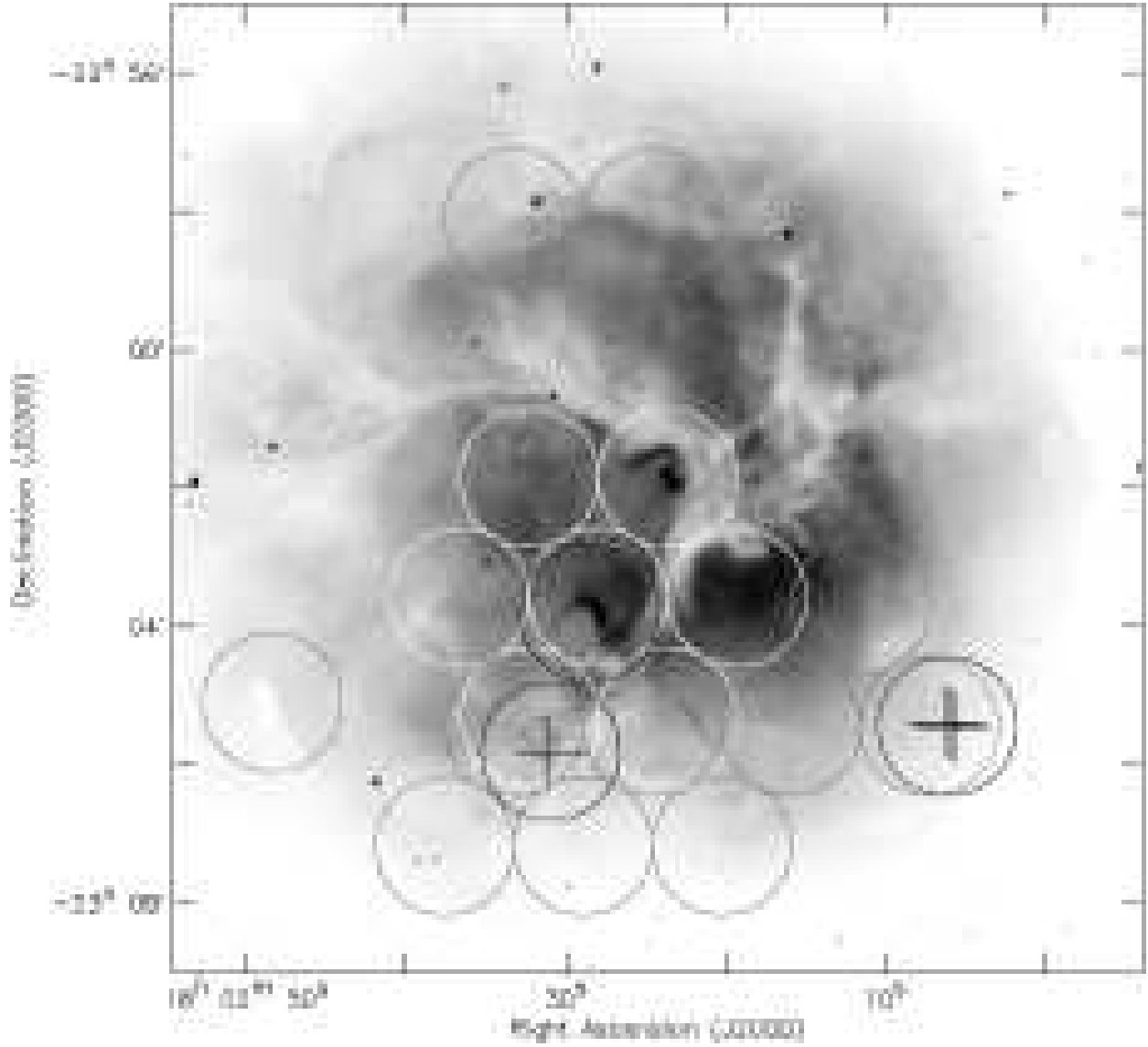}
\caption{\label{m20finder} Pointing map and maser locations for M20.
Caption as in Fig.~\ref{m16finder}.}
\end{figure}

\begin{figure}
\plotone{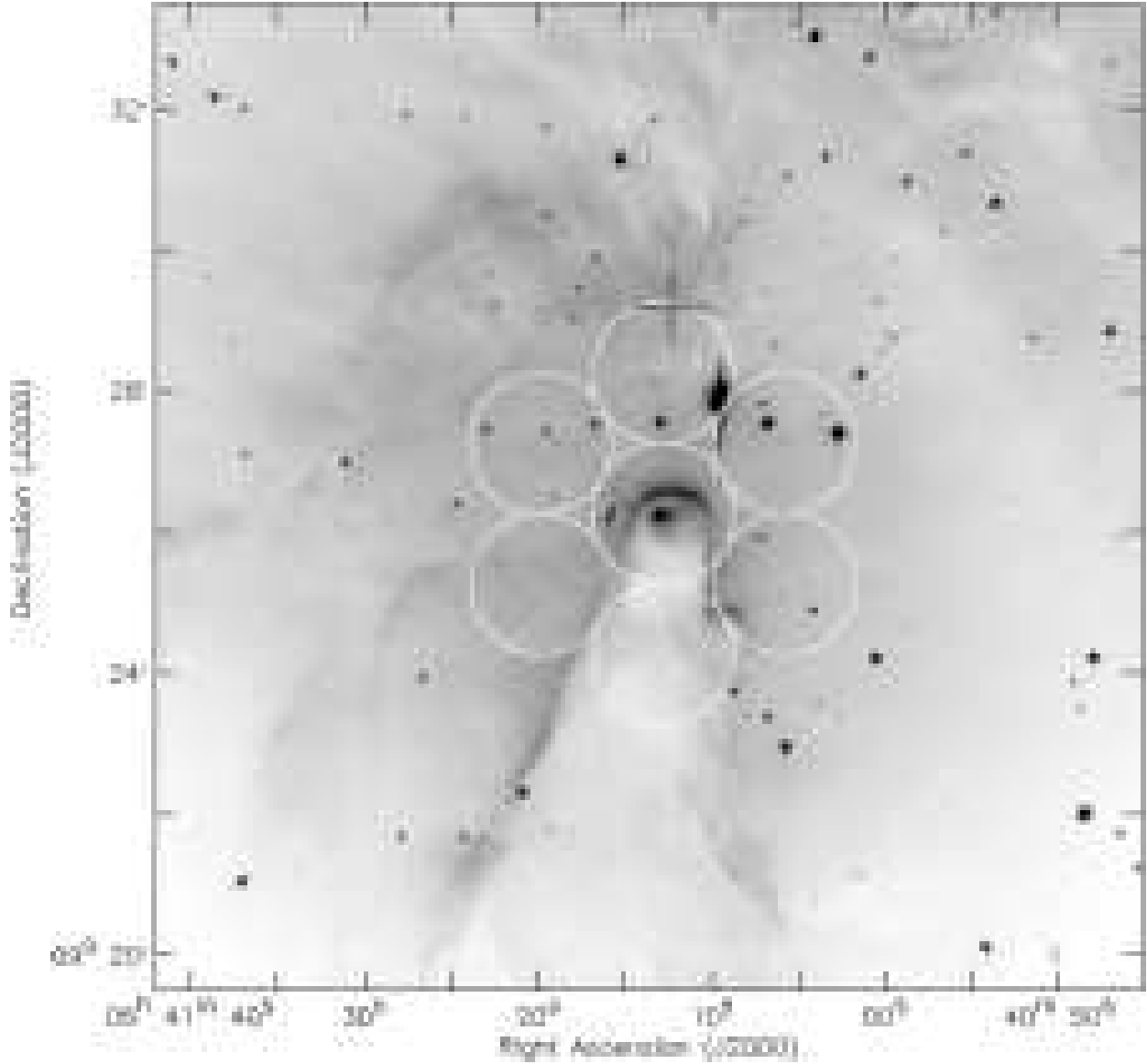}
\caption{\label{conefinder} Pointing map and maser locations for NGC~2264.
Caption as in Fig.\ref{m16finder}.}
\end{figure}

\begin{figure}
\plotone{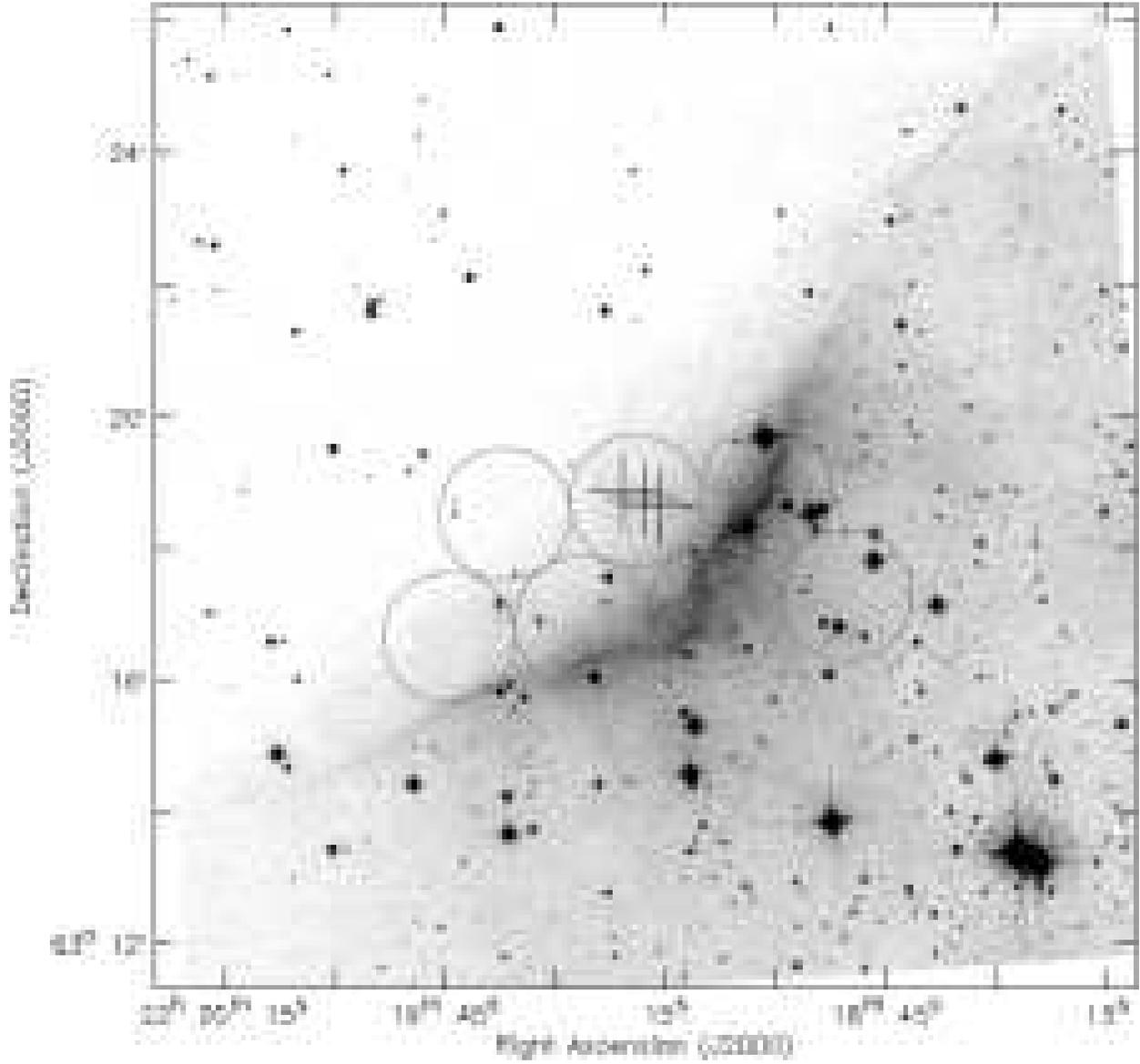}
\caption{\label{s140finder} Pointing map and maser locations for S140.
Caption as in Fig.~\ref{m16finder}, except greyscale image is extracted
from the Digital Sky Survey.}
\end{figure}

\begin{figure}
\plotone{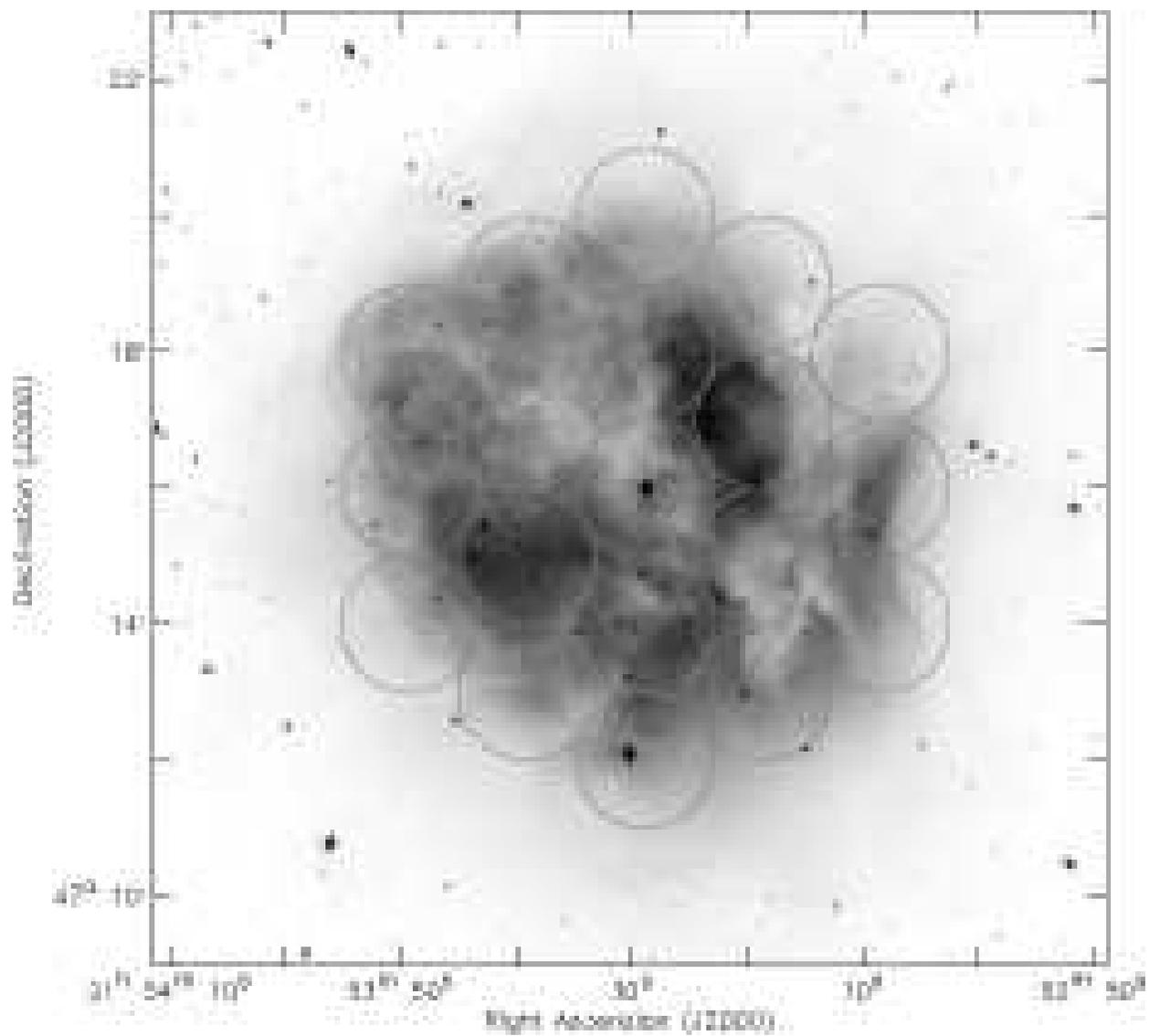}
\caption{\label{s125finder} Pointing map for S125. Caption as in
Fig.~\ref{m16finder}.}
\end{figure}

\begin{figure}
\plotone{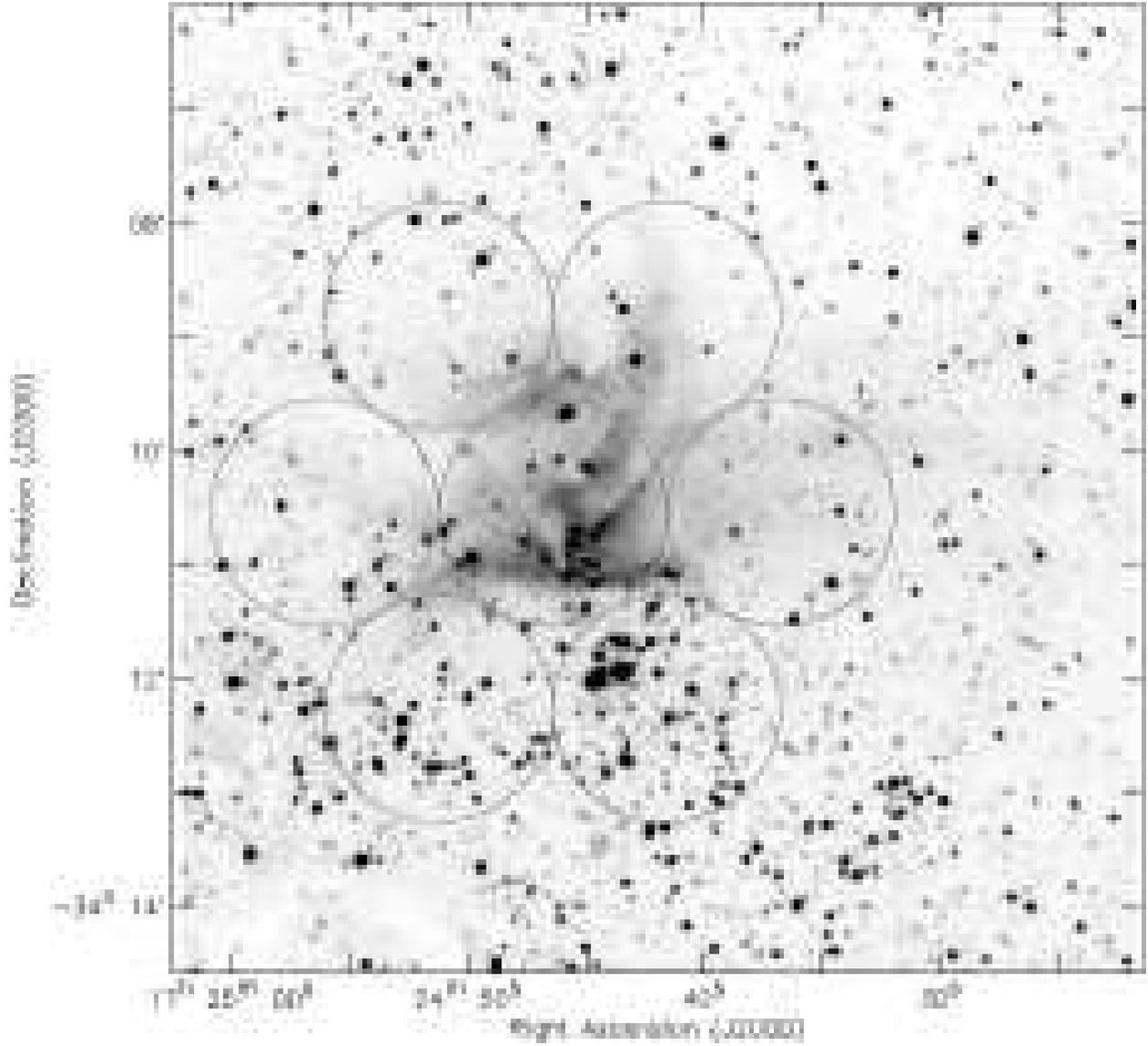}
\caption{\label{n6357finder} Pointing map for NGC~6357. Caption as in
Fig.~\ref{m16finder}, except greyscale image is extracted from the 2MASS
survey (K band).}
\end{figure}

\begin{figure}
\epsscale{1.0}
\plotone{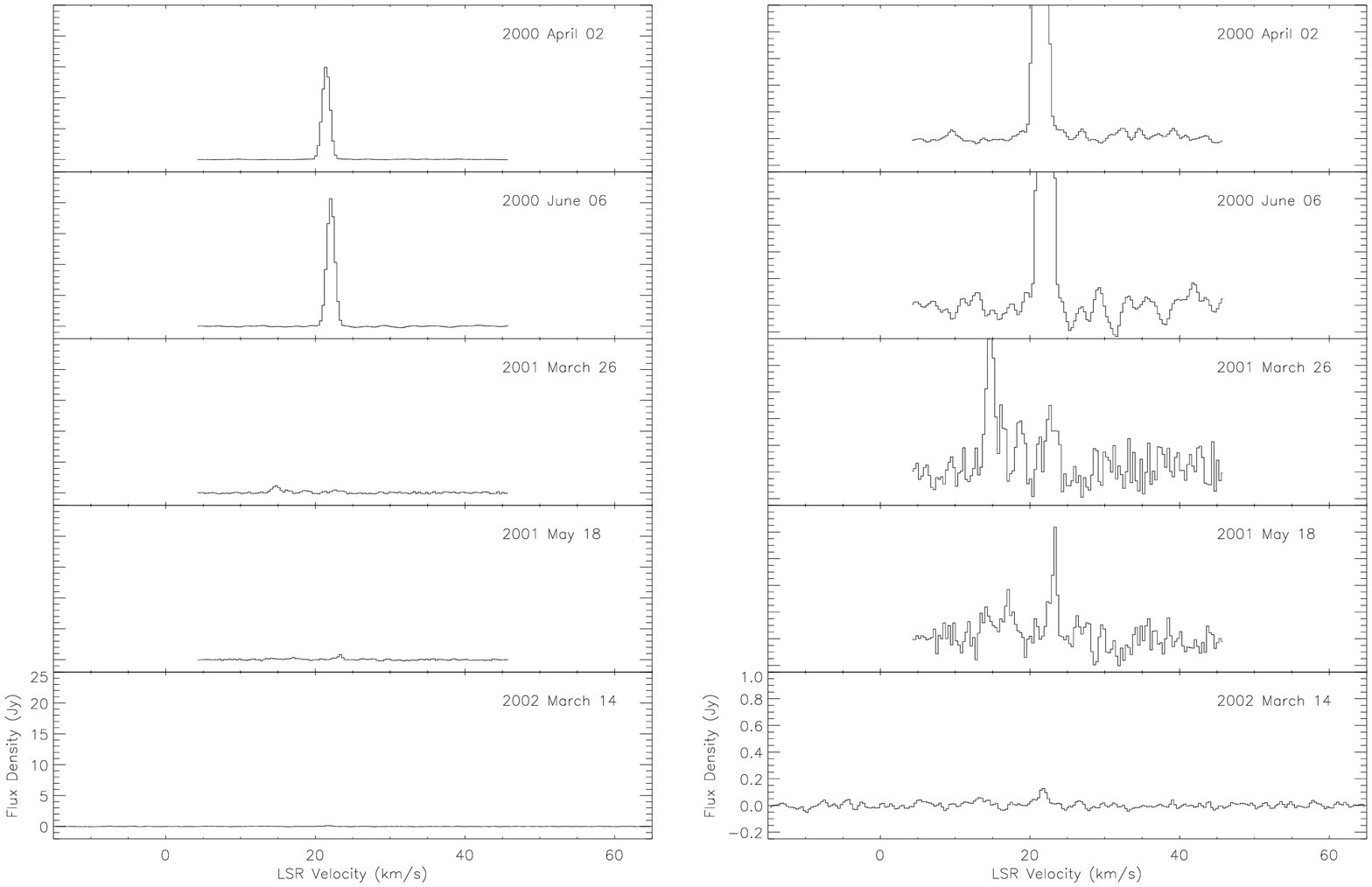}
\caption{\label{col2specs} Spectra of the M16 Column 2 maser. Dates of
observation are (top to bottom): 2000 April 02, 2000 June 06, 2001 March 26,
2001 May 18, and 2002 March 14. The left column of spectra are shown with a
common vertical scale that accomodates the brightest features of 2000, while
the right column of spectra are shown with a vertical scale that shows the
weaker features.}
\end{figure}

\begin{figure}
\epsscale{1.0}
\plotone{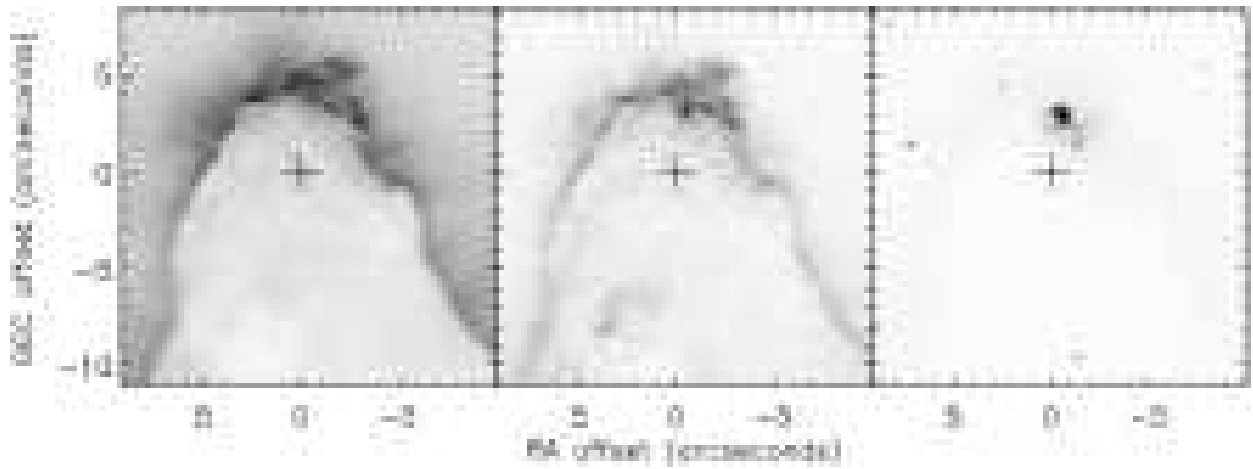}
\caption{\label{col2wfnic} HST WFPC2 and NICMOS images of the tip of
Column 2 in M16 showing the position of water maser: (left) H$\alpha$
(middle) \Stwo\ (right) F110. The water maser lies 2$\farcs$7
(P.A. $\sim$ 113$\degr$) from the source ES-2. The position of the
water maser on 2000 April 02 is marked by the black cross.}
\end{figure}

\begin{figure}
\plotone{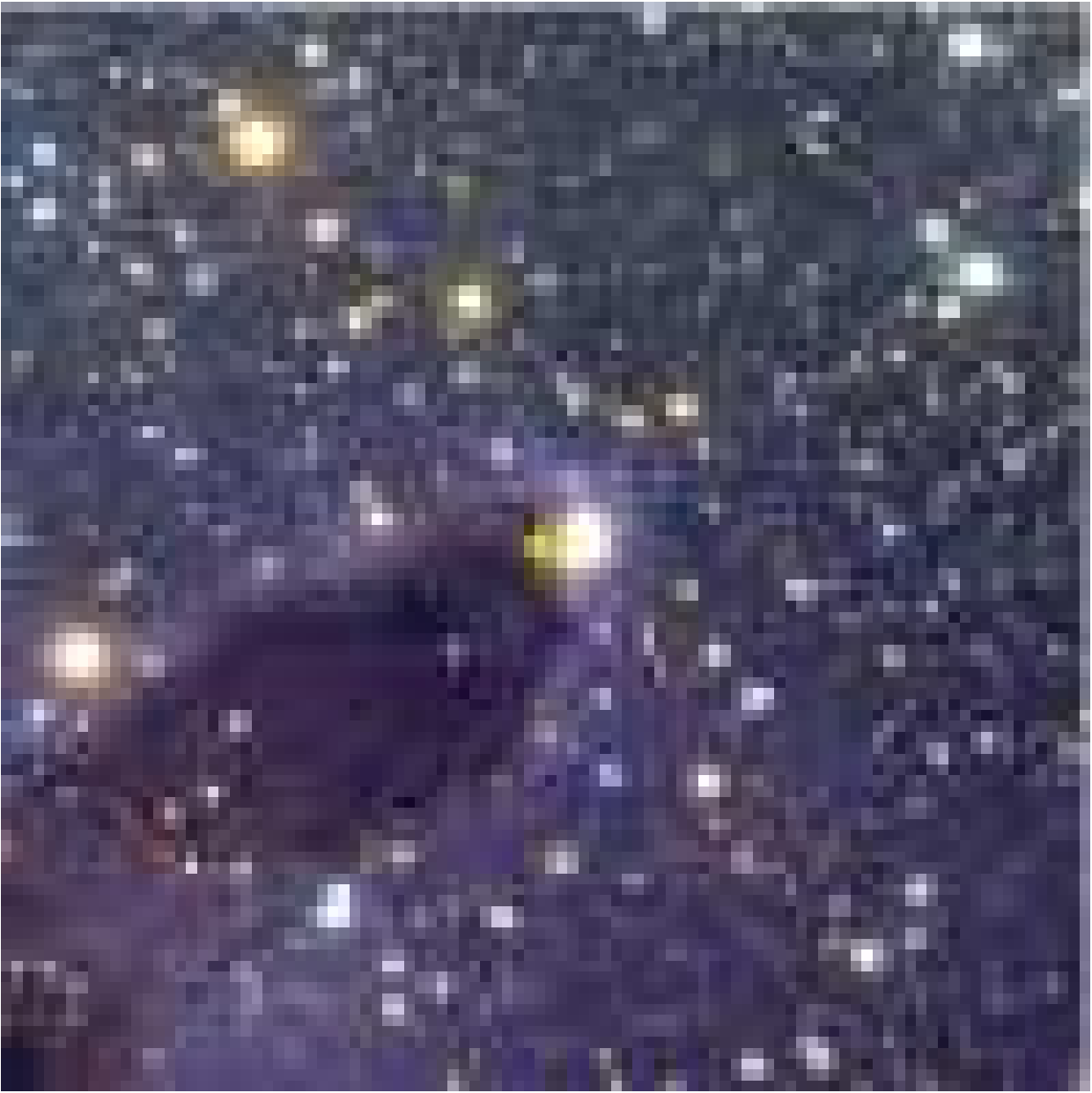}
\caption{\label{col2jhk} VLT ISAAC JHK 3-color composite of Column 2
from \citet{ma01}. Positions of detected water masers over three years
of observation shown (yellow crosses).}
\end{figure}

\begin{figure}
\epsscale{1.0}
\plotone{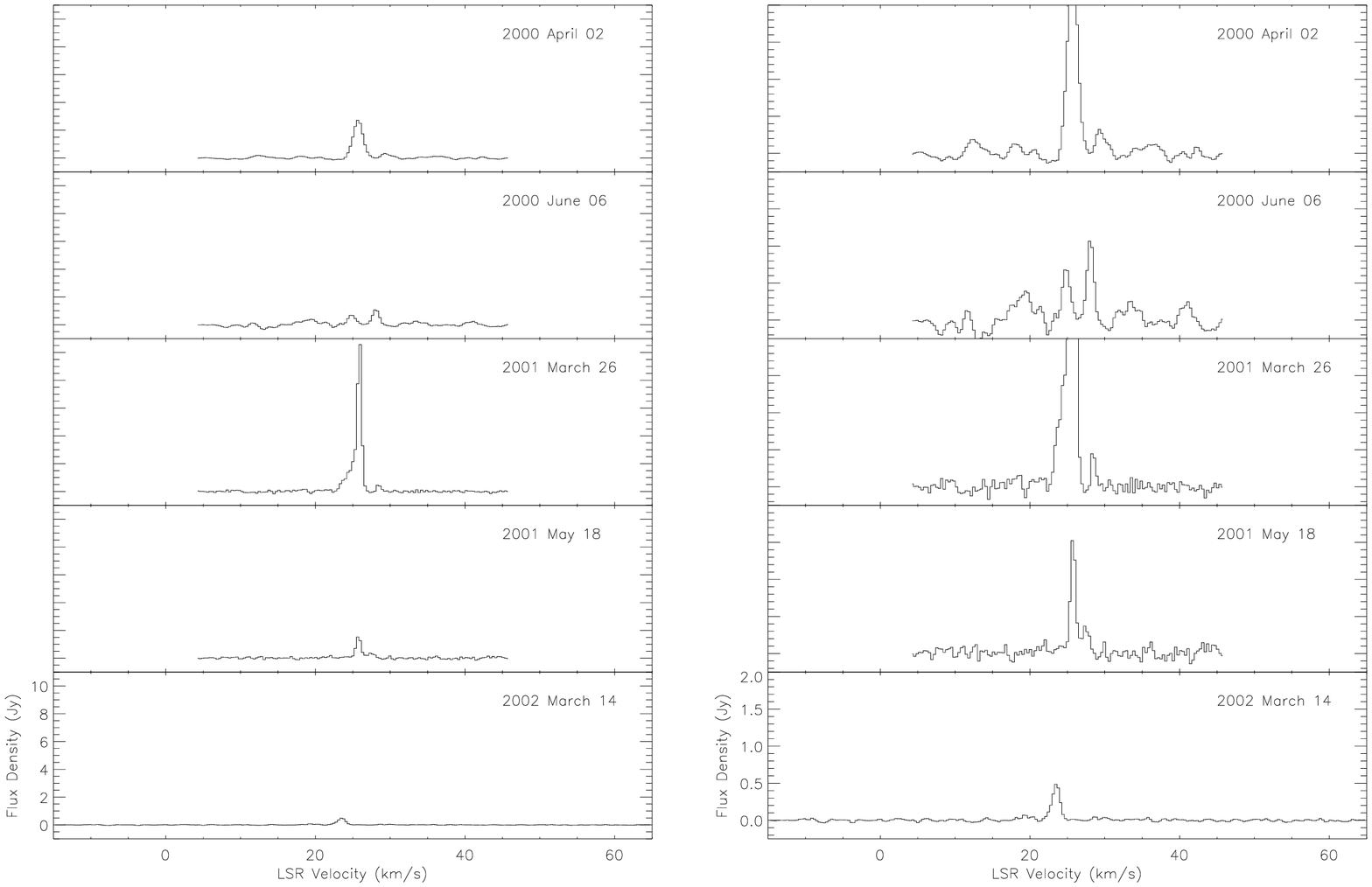}
\caption{\label{col4specs} Spectra of the M16 Column 4 maser. Dates of
observation are (top to bottom): 2000 April 02, 2000 June 06, 2001 March 26,
2001 May 18, and 2002 March 14. The left column of spectra are shown with a
common vertical scale that accomodates the brightest feature of 2001, while
the right column of spectra are shown with a vertical scale that shows the
weaker features.}
\end{figure}

\begin{figure}
\plotone{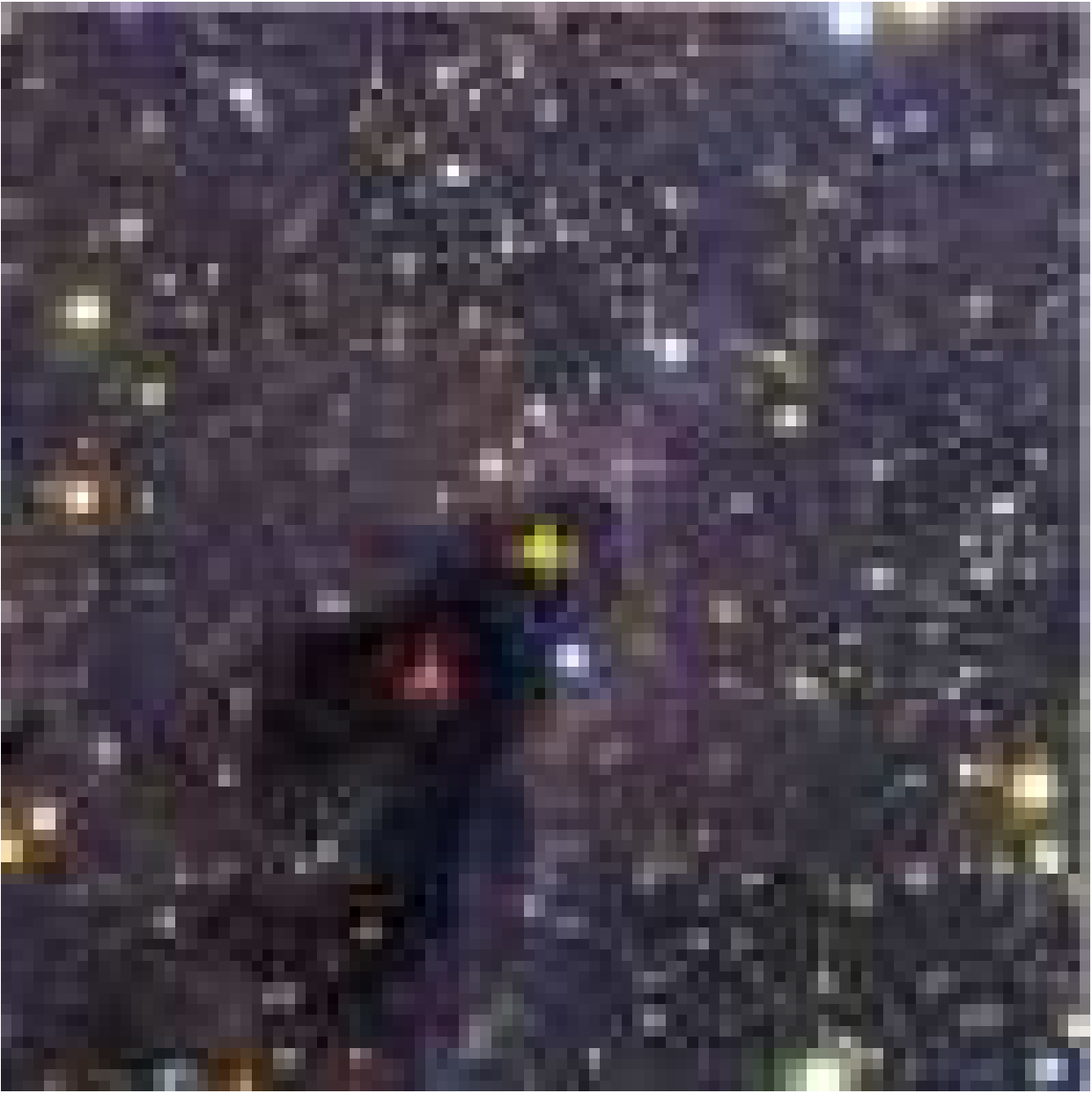}
\caption{\label{col4jhk} VLT ISAAC JHK 3-color composite of Column 4
from \citet{ma01}. Caption as in Fig.~\ref{col2jhk}.}
\end{figure}

\begin{figure}
\epsscale{0.6}
\plotone{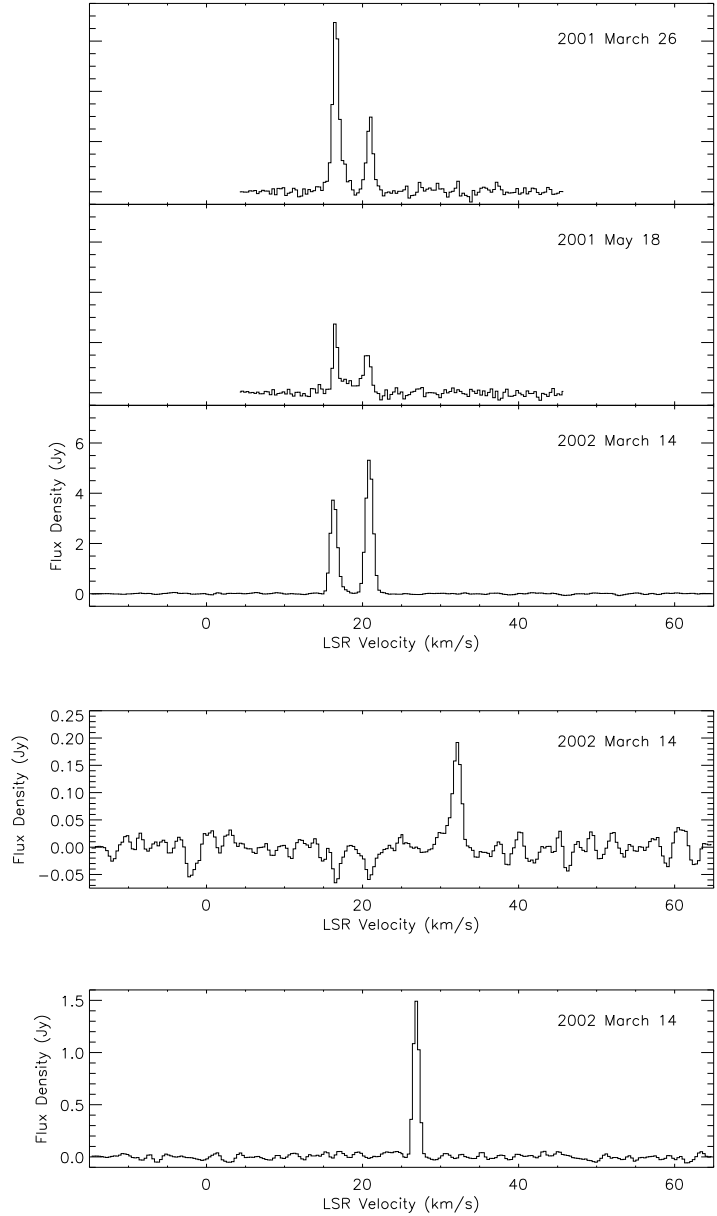}
\caption{\label{col5specs} Spectra of the M16 Column 5 maser
components A, B, and C. (Top) Spectra of component A on 2001 March 26,
2001 May 18, and 2002 March 14. (Middle) Spectrum of component B on
2002 March 14. (Bottom) Spectrum of component C on 2002 March 14.}
\end{figure}

\begin{figure}
\epsscale{1.0}
\plotone{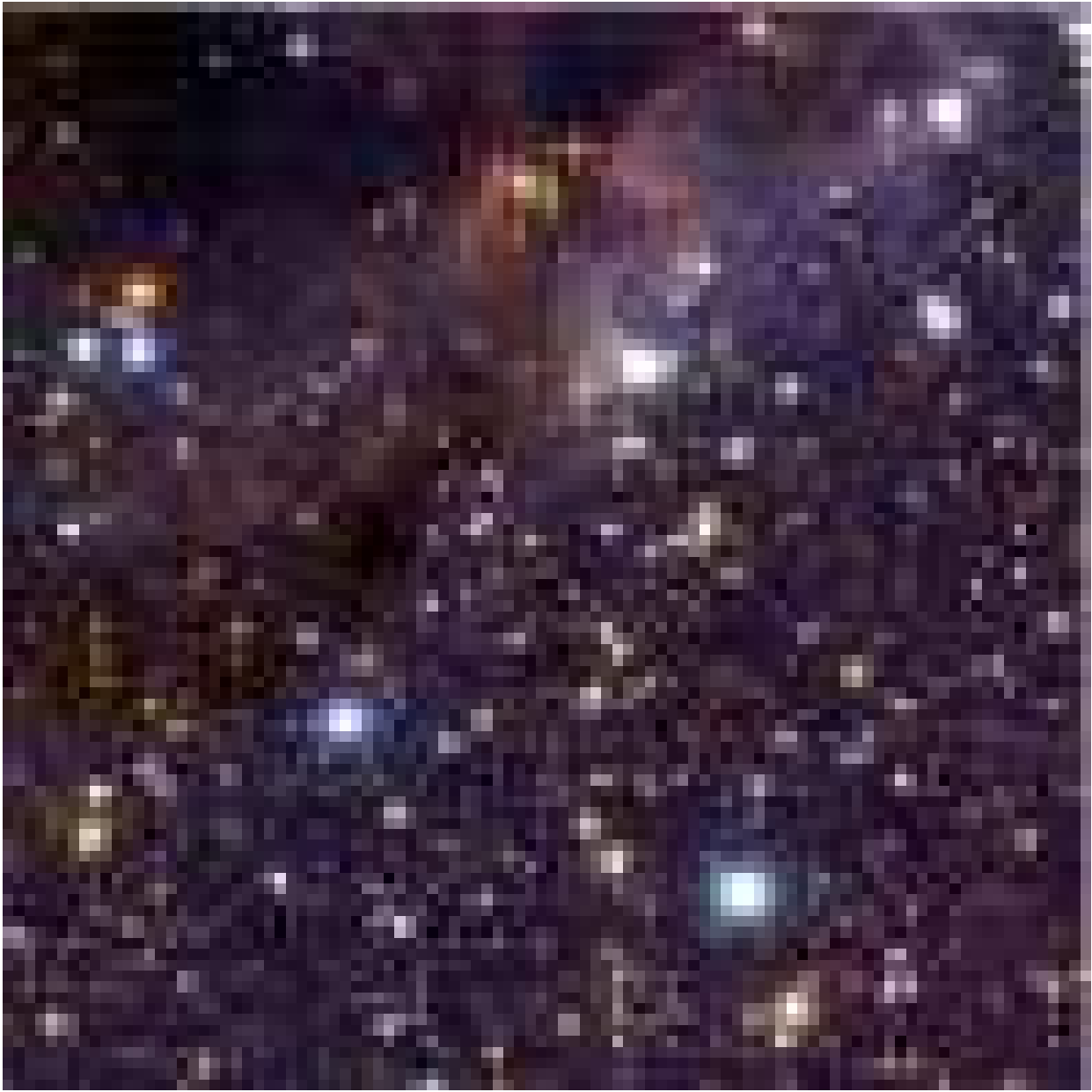}
\caption{\label{col5jhk} VLT ISAAC JHK 3-color composite of Column 5
from \citet{ma01}. Caption as in Fig.~\ref{col2jhk}.}
\end{figure}

\begin{figure}
\epsscale{0.7}
\plotone{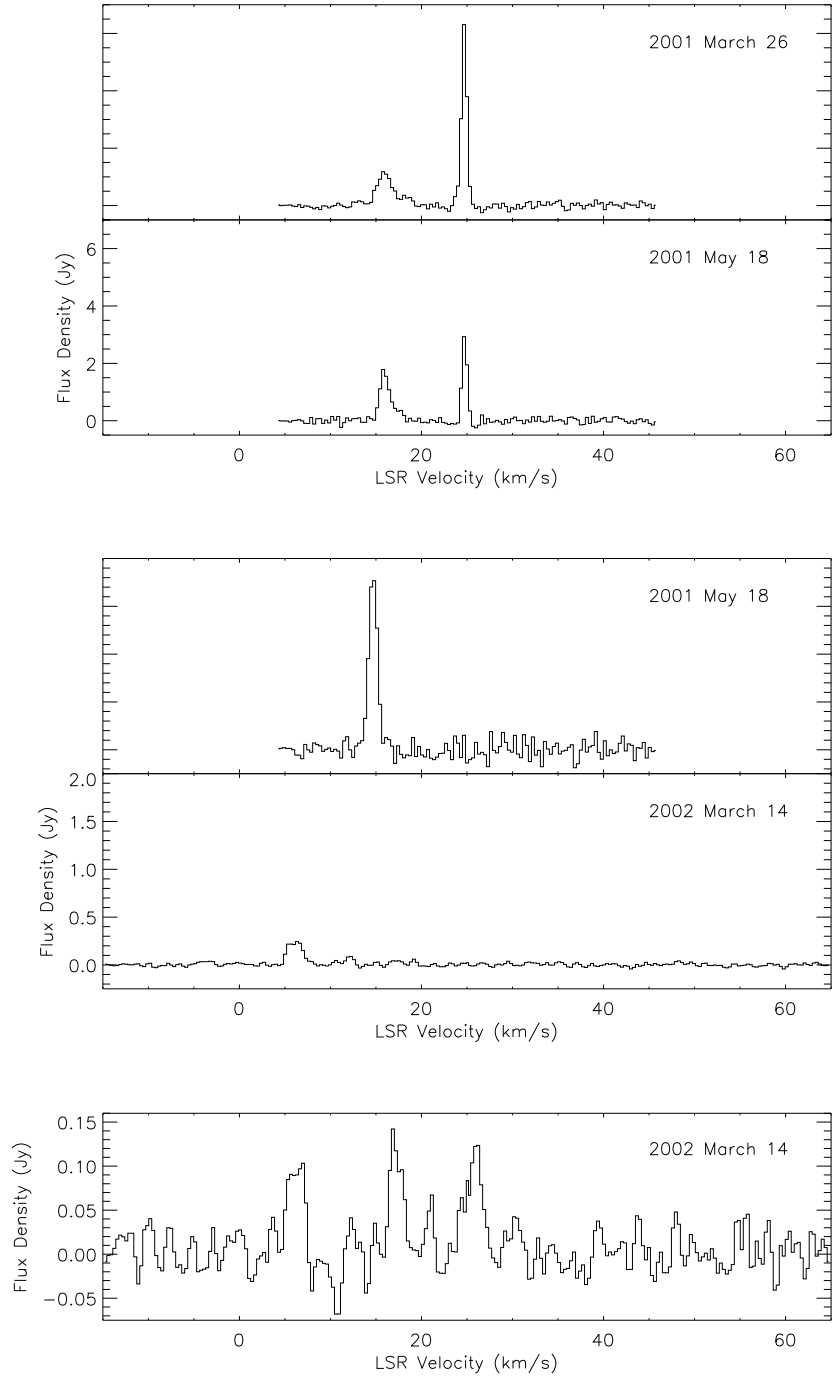}
\caption{\label{nbayspecs} Spectra of the M16 North Bay maser
components A, B, and C. (Top) Spectra of component A on 2001 March 26
and 2001 May 18. (Middle) Spectra of component B on 2001 May 18 and
2002 March 14. (Bottom) Spectrum of component C on 2002 March 14.}
\end{figure}

Previous searches for water maser activity in M20 do not appear in the
literature. The youth and intensity of star formation activity in M20
makes this \Htwo\ region an obvious choice for our survey. We focused
our attention on the south-central portion of M20 during the three
years of observations (see Figure~\ref{m20finder}). We targeted TC2 in
all three years to search for water maser activity powered by the
embedded source of the HH399 jet, but none was detected to a
sensitivity limit of 0.1 Jy (3$\sigma$).

\begin{figure}
\epsscale{1.0}
\plotone{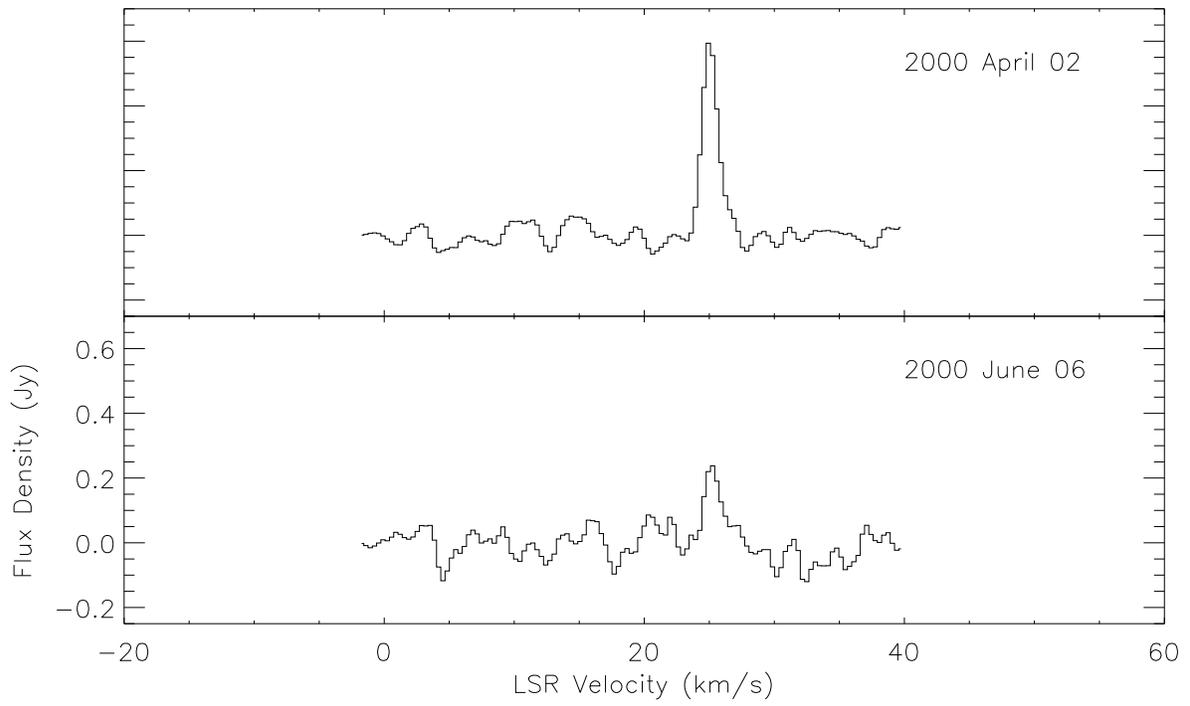}
\caption{\label{tsespecs} Spectra of the M20 southeast maser. Dates of
observations are 2000 April 02 (top) and 2000 June 06 (bottom).}
\end{figure}

In our 2000 April 02 and June 06 observations, we detected water maser
emission about 1$\arcmin$ south of TC2 or about 5$\arcmin$ south of
the center of M20. This maser has a single velocity component at a
\vlsr\ around +25 \kms\ (Figure~\ref{tsespecs}).  Emission was not
detected in our 2001 or 2002 epochs. Figure~\ref{m20finder} shows the
position of this maser within M20. In the infrared and millimeter maps
of \citet{cer98}, there is a spatially broad peak at the position of
the water maser. There is no visible star or optical signature of
outflows present at this position. The 2MASS point source
1802309-230549 lies 0$\farcs$6 from the position of the maser. This
source has {\it JHK} colors that imply it is embedded (i.e. A$_V$
$\sim$ 27-30) and luminous (i.e. $\sim$1500-2000 \Lsun), suggestive of
an extremely young low-mass or an intermediate-mass protostar.  The
projected separation between the maser and the 2MASS source
($\sim$1000 AU) is only somewhat larger than the typical separation
\citep[see, e.g.][]{che95,cla98,fur00} between a protostar and the
masers it excites, so this maser might either be associated with 2MASS
1802309-23054, or with a close neighbor of that source. If the maser
is associated with the 2MASS source, the visibility of the source in
the near-infrared makes it likely this is a Class I rather than a
Class 0 protostar.

\begin{figure}
\epsscale{1.0}
\plotone{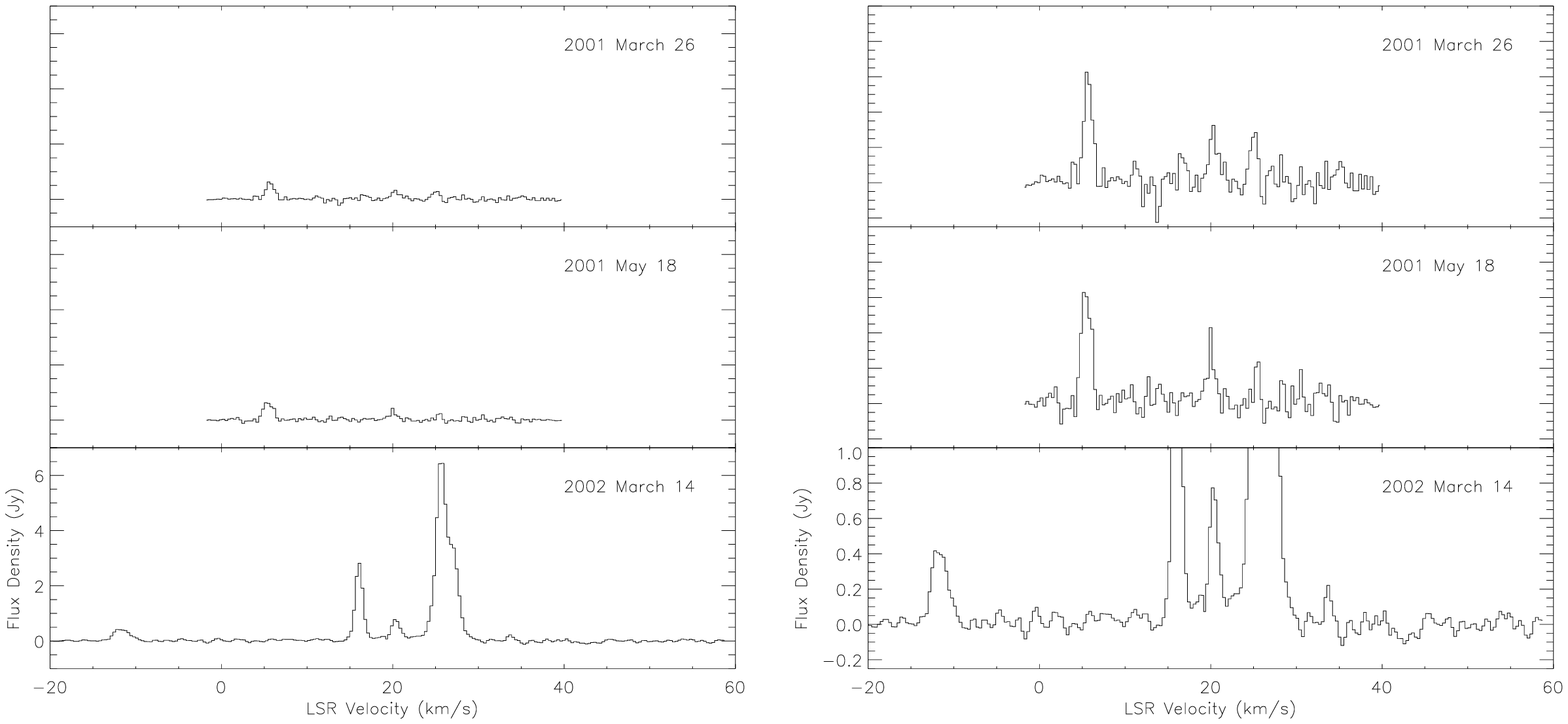}
\caption{\label{tc3aspecs} Spectra of the M20 TC3 maser component A.
Dates of observation are (top to bottom): 2001 March 26, 2001 May 18,
and 2002 March 14. The left column of spectra are shown with a common
vertical scale that accomodates the brightest feature of 2002, while
the right column of spectra are shown with a vertical scale that shows
the weaker features.}
\end{figure}

\begin{figure}
\epsscale{1.0}
\plotone{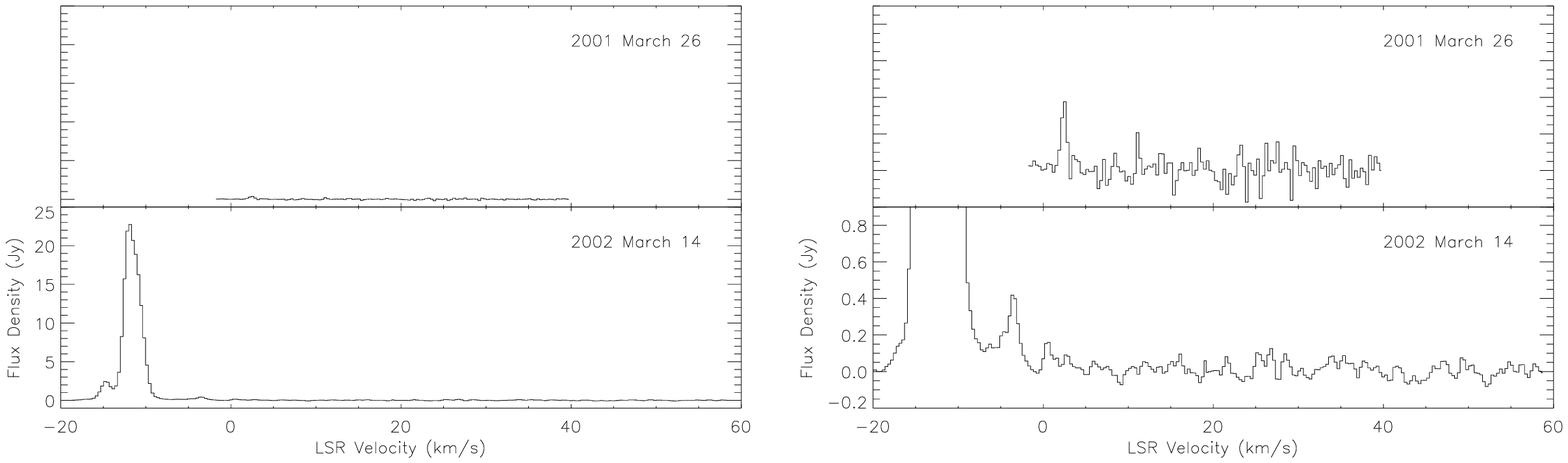}
\caption{\label{tc3bspecs} Spectra of the M20 TC3 maser component B.
Dates of observation are (top to bottom): 2001 March 26 and 2002 March 14.
The left column of spectra are shown with a common vertical scale that
accomodates the brightest feature of 2002, while the right column of spectra
are shown with a vertical scale that shows the weaker features.}
\end{figure}

\begin{figure}
\epsscale{0.75}
\plotone{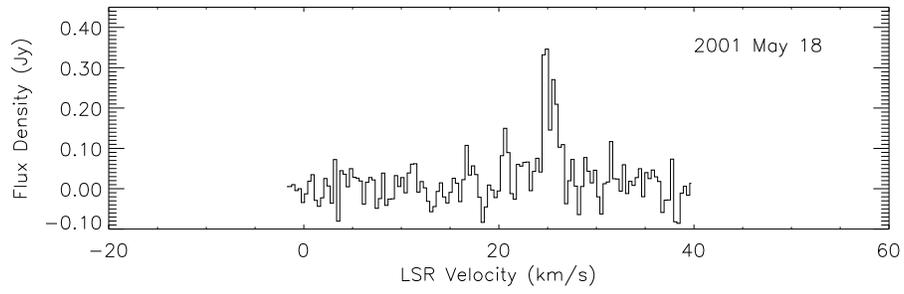}
\caption{\label{tc3cspec} Spectrum of the M20 TC3 maser component C on
2001 May 18.}
\end{figure}

In our 2001 and 2002 observations, we detected three new masers in the
southwestern portion of M20 very near the peak of the millimeter
source TC3 \citep{cer98}. Figures~\ref{tc3aspecs} - \ref{tc3cspec}
show the spectra of these masers. Component A shows multiple velocity
features, with the features around \vlsr\ = +20 \kms\ and +25.5 \kms\
present in all three epochs from 2001 March 26 to 2002 March
14. Component B shows a single weak feature in 2001 March 26 and two
strong features detected in 2002 March 14. The 22.7 Jy feature at
\vlsr\ = $-$11.8 \kms\ was the strongest emission we detected in any
of the \Htwo\ regions we observed. Component C was detected only on
2001 May 18 and consisted of a single weak feature at \vlsr\ = +24.9
\kms.

Figure~\ref{m20finder} shows the positions of these three new masers.
All components are within 5$\arcsec$ from the peak of 1250 $\micron$
continuum emission of TC3 and separated from each other by
$\sim$5$\arcsec$. Assuming a distance of 1.68 kpc to M20, this
separation corresponds to a physical distance of 0.04 pc.
\citet{lc00} calculated a Jeans length of 0.06 pc in the core of TC3,
comparable to the separation between the observed water masers. Thus,
it is likely that multiple low-mass protostars have formed within
TC3. If the detection rate of water maser activity from low-luminosity
protostars is about one-third \citep{cla96}, then we expect $\sim$9
protostars to be present within the core of TC3. The maps presented by
\citet{lc00} do not have the spatial resolution required to detect
substructure on this scale within TC3. \citet{lc00} reported the
detection of outflow signatures in SiO and CS lines, but the mass (1.1
\Msun) and luminosity (1.2 \Lsun) inferred from the molecular lines
are low enough to be attributable to many young, low-mass protostars
contributing to an unresolved outflow signature.

\subsection{NGC~2264}

At the southern end of the cluster and \Htwo\ region NGC~2264 lies the
Cone Nebula. The Cone is illuminated by the stars HD 47887 and 47887B
of spectral type B2III and B9V, respectively. The stellar population
in this region shows \Ha\ and near-infrared excesses indicative of
youth \citep{lyg93,sbl97}. Roughly 3$\arcmin$ north of the Cone is a
luminous infrared source IRS~1 that powers a bipolar outflow. This
portion of the \Htwo\ region has been mapped in several molecular
transitions by \citet{sch97} and in submillimeter continuum by
\citet{war00}. The distance to NGC~2264 has been suggested to be about
800 pc \citep{sj83}.

\begin{figure}
\epsscale{0.75}
\plotone{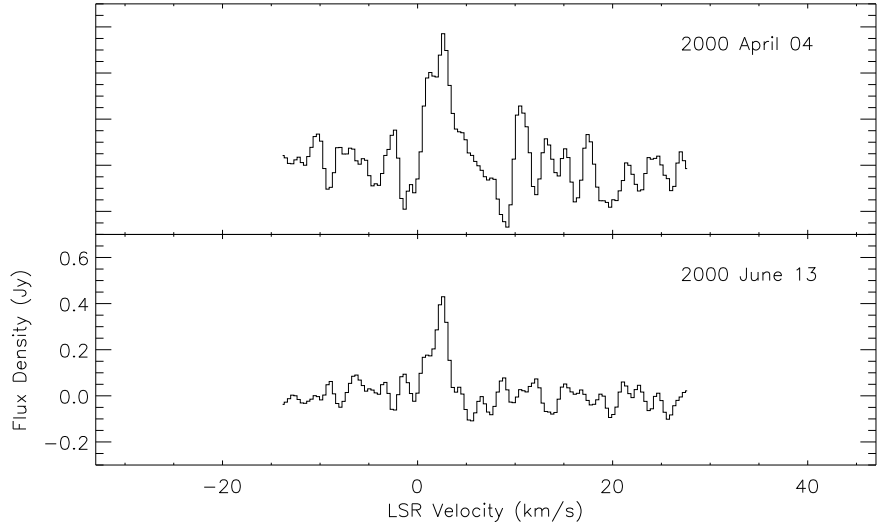}
\caption{\label{conespecs} Spectra of the Cone maser on 2000 April 04
(top) and 2000 June 13 (bottom).}
\end{figure}

On 2000 April 04 and June 13, we searched a portion of the \Htwo\
region centered on the tip of the Cone Nebula
(Figure~\ref{conefinder}).  Water maser emission associated with
NGC~2264 IRS~1 is well-known and well characterized as highly variable
\citep{gd77,tof95}.  One of our pointings contained the position of
IRS~1. We detected no water maser emission from the position of IRS~1
to a limit of 0.2 Jy (3$\sigma$). We detected a previously unreported
water maser on 2000 June 13, at a position approximately 45$\arcsec$
southeast of IRS~1 at \vlsr\ = +2.6 \kms. The 2000 April 04
observation of the same field contained a weak (5.9$\sigma$) component
at the same position and velocity as the June 13 detection.
Figure~\ref{conespecs} shows the spectra from both dates of
observation. The positions, peak fluxes, and velocities of the peak
emission for this maser are listed in Table~\ref{detected}.

This water maser lies within 0.3$\arcsec$ (240 AU at 800 pc) of the
center of MMS3, a warm, dense core detected in submillimeter continuum
observations by \citet{war00}. The derived mass for MMS3 from the dust
continuum emission is 48 \Msun.  \citet{sch97} detected a bipolar
outflow at this same position in their CS 5$\rightarrow$4 map.
\citet{war00} postulated that MMS3 had formed a massive Class 0 object
at its center. The detection of water maser activity from this site
appears to confirm this conclusion.

\subsection{S140}

Sharpless 140 \citep[S140,][]{sha59} is a bow-shaped \Htwo\ region
powered by the B0 star HD 211880 on the southeast edge of the L1204
dark cloud. The interface between the \Htwo\ region and the adjoining
molecular cloud is viewed mostly edge-on. This region shows signs of
ongoing star formation, the clearest evidence being a cluster of
luminous infrared sources, designated IRS~1-3, which lie within the
molecular cloud about 1$\arcmin$ behind the \Htwo\ region interface.
These sources have been studied extensively at infrared wavelengths
\citep[for recent high resolution {\it K}-band observations,
see][]{pre01,ps02,wei02}. These observations highlight intense
outflow activity originating from IRS~1 and IRS~3.  Optical observations
by \citet{bal02} show strong outflow activity from numerous sources in
the region, suggesting further sources of outflows. S140 is estimated
to be at a distance of 900 pc \citep{gg76}.

Well-known water maser activity is associated with IRS~1
\citep{km76,gd79,tof95}.  On 2000 June 06, we searched a portion of
the \Htwo\ region as well as the molecular cloud in S140 (see
Figure~\ref{s140finder}).  We detected 3 masers, which we designate
S140 component A (associated with IRS~1), B (to the SW of IRS~1), and
C (to the NE of IRS~1). The positions, peak fluxes, and peak
velocities are listed in Table~\ref{detected}.  Component A contains
two spatially unresolved velocity components at \vlsr\ = $-$15.3 \kms\
and +3.5 \kms. \citet{tof95} resolved component A into velocity
components at \vlsr\ of $-$18 \kms\ and +4.1 \kms\ and separated by
$\sim0\farcs5$ from each other and separated by $0\farcs1$ and
$0\farcs4$ from IRS~1, respectively. Our detected Component A is most
likely excited by IRS~1.

\begin{figure}
\epsscale{0.6}
\plotone{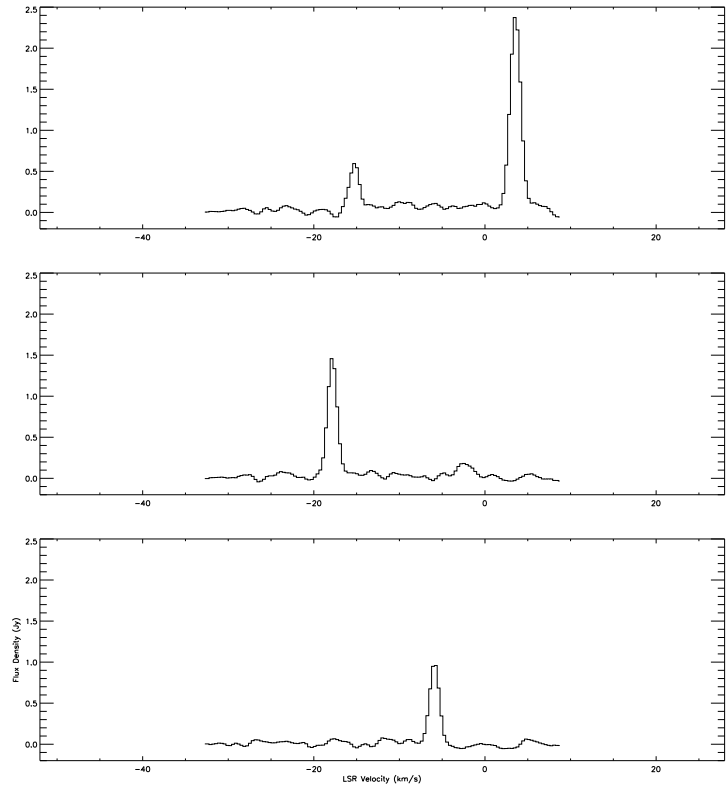}
\caption{\label{s140specs} Spectra of the S140 masers observed on
2000 June 06 (top to bottom): components A, B, and C.}
\end{figure}

In addition to the well-known water maser activity associated with
IRS~1, we have detected components B and C at \vlsr\ of $-$17.9 and
$-$5.7 \kms, respectively (see Figure~\ref{s140specs}). These velocity
components appear to have been detected previously in single-dish
observations, but our observations are the first to spatially resolve
components B and C from component A \citep[see, Fig. 27 and 28
in][]{val02}. No infrared sources appear at the positions of
components B and C in the images of \citet{pre01}.  However, these
components are aligned with the strong outflow from IRS~3
\citep{pre01,ps02,bal02}, and may be excited either by that outflow or
by additional protostars in the region.

\subsection{S125}

Sharpless 125 \citep[S125,][]{sha59} is a classic \Htwo\ region
symmetrically distributed around and powered by the early B-type star
BD $+46\degr3474$. \Hone\ observations by \citet{ri82} confirm the
basic picture of a blister \Htwo\ region formed about $10^5$ years ago
by the breakout of ionized gas heated by BD $+46\degr3474$. Optical
observations of the stellar population by \citet{hd02} put the age of
the cluster at 1 Myr with a large uncertainty. The photometry of
\citet{hd02} implies a distance to S125 of 1.2 kpc.

We included S125 in our survey as an example of a low-ionization
\Htwo\ region. In their search for water maser emission near OB
associations, \citet{bl79} found no water masers in this region to a
sensitivity of 6 Jy (3$\sigma$). We surveyed the extent of the
optically bright portion of S125 on 2001 March 26 and May 18 using 19
pointings symmetrically arrayed around BD $+46\degr3474$ (see
Figure~\ref{s125finder}).  We detected no water maser emission on
either date to a sensitivity of 0.1 Jy (3$\sigma$).  While there
exists a population of low-mass stars in S125, \citet{hd02} argue that
most of these objects have ages of around $10^6$ years.  The lack of
maser emission in S125 supports the conclusion that there are few
young protostars in S125, and that star formation in the region ended
some time ago.

\subsection{NGC~6357}

The NGC~6357 star forming complex includes several distinct \Htwo\
regions in different stages of evolution, with G353.2+0.9 being the
youngest and brightest. In their study of NGC~6357, \citet{fel90}
found G353.2+0.9 is characterized by many signatures of star
formation, such as strong H$\alpha$ and radio continuum emission and
multiple near- and mid-infrared sources. Their optical images of
G353.2+0.9 also shows the presence of an ``elephant trunk'' pointing
at the cluster C1722$-$343 (Pismis 24), which lies about 30$\arcsec$
to the south of G353.2+0.9 \citep{pis59}. The cluster C1722$-$343
contains several O stars, the brightest being HD 319718 which has been
variously classified as O4 to O7 \citep{ltn84,nec84}. The combined
ionizing flux from C1722$-$343 is enough to explain the high
excitation lines observed by \citet{ltn84}. The photometry of
\citet{ltn84} and \citet{nec84} agrees with previous distance
estimates to NGC~6357 of 1.7 kpc.

We included G353.2+0.9 in our survey due to the apparent close
proximity of the ionizing cluster to the molecular cloud \citep[see][for
maps of molecular line emission]{mbf97}. The implied high ionizing
flux would likely cause rapid heating and photoevaporation of the star
forming sites within G353.2+0.9. The disruption of the star-forming
environment could lead to a strong observable effect on the young
stellar objects within the immediate vicinity of the ionization front.

Water maser activity has been detected within NGC~6357, but the
nearest water maser to G353.2+0.9 lies $\sim$0.3$\degr$ away
\citep{sak84}.  The sensitivity limit of that survey was slightly less
than 1 Jy.  We surveyed the extent of the optically bright portion of
G353.2+0.9 on 2001 March 26 and May 18 using 7 pointings symmetrically
arrayed around the base of the elephant trunk feature (see
Figure~\ref{n6357finder}). We detected no water maser emission from
this region on either date to a sensitivity of 0.1 Jy
(3$\sigma$). There are strong indications from radio, infrared, and
optical observations that this region harbors active star formation
\citep{fel90}, so the absence of water masers is not due to the
absence of young protostars. Rather, as discussed below, it seems
likely that the water maser-exciting phase of the evolution of a
protostar is truncated in the \Htwo\ region environment.

\section{Discussion}

Before proceeding with our discussion, it is worth making the point
that we have found water maser emission to be a useful tracer of the
youngest stages of star formation in \Htwo\ region environments. The
water masers in NGC~2264 and S140 appear to be excited by massive
protostars.  We have provided arguments in support of our conclusion
that the maser-exciting protostars in M16 and M20 are low mass.
However, the use of water masers as signposts of protostars does not
depend on the accuracy of our mass determinations. We were successful
in both clarifying the star formation present at previously known
locations, as well as identifying new sites of star formation in these
complex regions.  We are optimistic about the future application of
this technique.

Discussion of this initial survey of water maser emission in \Htwo\
region environments will focus on two basic questions.  The first is
the implication of our results with respect to triggering of star
formation in such environments (Section 4.1).  The second is the
difference between water maser activity in \Htwo\ regions and regions
of isolated star formation, and the implications for how the
environments of evolving protostars are affected (Section 4.2).  Each
of these will be considered in light of the scenario for star
formation presented by \citet{hes96}.

\subsection{Evidence for Triggering of Star Formation}

The spatial distribution of water masers found in this study suggest
that low-mass star formation in \Htwo\ region environments is
concentrated in molecular gas that is located very close to the
current location of the ionization front.  That gas has been
compressed by the ``ionization shock'' that is driven in advance of
the ionization front.  This can be most clearly seen in M16 where the
silhouette geometry of the columns means there is less ambiguity in
the line-of-sight position of protostars relative to ionization
fronts. As can be seen in Figure~\ref{m16finder}, all masers we have
detected in M16 lie a short distance from ionization fronts.  Only
12\% of the field surveyed has a projected distance of less than 0.2
pc from an ionization front. Even so, 6 of the 8 water masers detected
lie in this zone. The odds of a population of protostars randomly
distributed throughout the surveyed area lying within the zone by
chance are less than $10^{-4}$. If, instead, the protostars are
randomly distributed throughout the columns only, the odds are less
than 3\%. In either case, the concentration of young protostars near
ionization fronts appears to be a real effect. \citet{whi99} present CO
observations of the gas in the columns, and argue that shocks with
$v_s \sim 1-2$ km s$^{-1}$ are currently being driven into the columns
of gas, which have only been exposed to intense UV for of order $10^5$
years.  \citet{fhs02} present CO observations at higher resolution,
and find that the arc-like shape of the densest molecular gas shows the
action of this advancing shock.  They go even further to claim that in
Column 1 they resolve a linear progression in the properties of
molecular gas and young stellar objects with increasing distance from
the ionization front.

The two masers that are not seen in projection within 0.2 pc of an
ionization front in M16 lie within the North Bay.  In this region, the
ionization front is more face-on, and is located on the back side of a
dark cloud seen in silhouette against the \Htwo\ region.  As such, the
maser positions are not constrained by a simple cylindrical geometry
as in the columns. These two masers may be within 0.2 pc of the
ionization front, but we are unable to tell.  They do, however, appear
to be located in gas that is interacting with the \Htwo\ region.  

From this analysis, we conclude that the distribution of water masers
in M 16 and the underlying young protostellar population is not
random, but is concentrated within a layer $\lesssim$0.2 pc from the
ionization front.  The localization of young protostars very near the
ionization front in M 16 is further supported by the results of
\citet{tsh02} who presented {\it HST} NICMOS observations of the
region.  Each of the bright protostars seen in those observations lie
extremely close to the ionization front.

The geometry of the other \Htwo\ regions we surveyed is not suitable
for a statistical analysis like that above, but the data are
suggestive.  The maser M20 SE lies in a region containing a number of
locations where the ionization front is tangential to the line of
sight, and is seen in projection about $10^{17}$ cm behind an
ionization front that is prominent in {\it HST} images of the region
\citep{hes99}.  The other maser-exciting protostars in M20 lie within
the TC3 molecular core, which \citet{cer98} and \citet{lc00} argue is
undergoing radiatively-driven implosion in advance of the ionization
front.

Taken as a whole, our data indicate that young protostars in \Htwo\
regions are concentrated in molecular gas that has been compressed by
ionization shocks.  This distribution provides strong circumstantial
evidence that most low-mass star formation in \Htwo\ regions is
triggered by those ionization shocks.  There is no reason to imagine,
in the absence of a causal connection, that the youngest protostars
would just happen to lie so close to ionization fronts.

\subsection{The Efficiency of Maser Production in \Htwo\ Regions is Low}

The regions that we observed in this survey were chosen in part
because of evidence of significant recent star formation.
\citet{hil93} describe M16 as a cluster ``caught in the act'' of star
formation -- a conclusion that is supported by other observers
\citep[e.g.][]{dew97,ma01}.  The presence of low-mass protostars
within the M16 EGGs has been discussed by \citet{ma01}, who find that
approximately 10 EGGs contain stellar and/or substellar objects that
can be seen at 2$\micron$.  More EGGs might contain far infrared
sources.  EGGs have lifetimes against photoevaporation of $\sim 10^4$
years \citep{hes96}, so the presence of at least 10 young stellar
objects within EGGs indicates that newly formed stars are emerging
into the interior of the \Htwo\ region at a rate of $\sim 10^{-3}$
stars year$^{-1}$.  If this represents star formation behind an
ionization shock that is still advancing along the columns, as argued
above, then this should be a good indication of the current rate of
star formation that is taking place behind the shock. This rate is in
line with the integrated star formation history of the cluster.
\citet{hil93} finds that the IMF in M 16 is close to Salpeter, with
141 stars with masses greater than 5 \Msun.  Integrating a Salpeter
IMF and assuming a cluster age of $2 \times 10^6$ years gives an
average star formation rate across the entire cluster of $2 \times
10^-2$ stars year$^{-1}$, or 20 times the star formation rate in the
columns inferred from the \citet{ma01} data.  This is also a
reasonable order-of-magnitude rate if star formation based on the
available gas and the rate at which the ionization shock is advancing
through the columns.  Observations in C$^{18}$O by \citet{whi99} show
about 200 \Msun\ of molecular gas in the columns.  A star formation
rate of $10^{-3}$ stars year$^{-1}$ is fully consistent with
converting this much gas into low-mass stars in a shock crossing time
of $\sim 10^5$ years and a star formation efficiency of a few tens of
percent.

A survey by \citet{vrc02} of the protostellar population in $\rho$ Oph
and Lynds dark clouds led those authors to conclude that the lifetime
of the Class 0 stage is $2\times10^5$ years.  A model developed by
\citet{ww01} similarly predicts a lifetime of almost $10^5$ years for
the Class 0 stage.  The correlation between the Class 0 stage of
protostar evolution and water maser activity \citep{fur01} thus
suggests that the typical maser-producing lifetime of an isolated
protostar is $\sim 10^5$ years.  If stars are forming within Columns
1, 2, and 3 at an ongoing rate of $10^{-3}$ stars year$^{-1}$, and
if each remains in a potentially maser-producing phase for $10^5$
years, then we would expect there to be about 100 protostars capable
of producing masers in these columns.  The fraction of such protostars
that actually exhibit water maser activity at any given time is about
30\% in the case of isolated low-mass star formation
\citep{cla96,fur01}, so we might expect to see several tens of water
masers in the M16 columns.  Instead we see one: M16 Column 2.  

Our observations of maser activity in other \Htwo\ regions bear out
our conclusion that water maser activity is less likely to be seen in
association with protostars in \Htwo\ region environments.  NGC~6357
in particular shows clear evidence of ongoing star formation near the
ionization front \citep{fel90}, but no water maser emission associated
with that star formation.  We tentatively conclude that a much smaller
fraction of young protostars excite water maser emission in \Htwo\
region environments than in regions of isolated star formation.  An
expanded, unbiased survey of water maser emission in \Htwo\ regions is
needed to strengthen this conclusion.

It should be noted that triggered star formation in \Htwo\ regions is
an ongoing process, and not a one-time event.  The spread of ages in
young stellar objects in \Htwo\ regions \citep[e.g.,][]{dew97,hil93}
is a natural consequence of this process.

\subsection{Disruption of Star Forming Environments by the Advancing
Ionization Front}

In the preceding paragraph, we find that water maser emission is less
likely to be seen in association with protostars in \Htwo\ region
environments.  Possible explanations for this effect can be broadly
divided into two groups: intrinsic properties of star-forming cores
and protostars in \Htwo\ regions, and external environmental effects
associated with the ongoing evolution of the \Htwo\ region. The first
group of explanations might address questions such as whether the
triggering mechanism leads to a shorter duration of the maser
excitation phase or a lower duty cycle of water maser activity.
Little is know about the detailed differences between the triggered
collapse of a protostar and the collapse of a protostar in isolation,
so there is little we can say about these possible explanations for
the low rate of maser activity.  Much more observational and
theoretical work is needed on the details of triggered collapse.

On the other hand, the evolution of the \Htwo\ region environment as a
whole is much easier to observe and much better characterized. In this
environment, the ionization shock triggers the collapse of cores, as
discussed above.  Some time later, these cores, and any protostars
they contain, will be overrun by the advancing photodissociation
region and then by the ionization front itself.  The space velocity of
a D-type ionization front trailing an ionization shock is very close
to the space velocity of the ionization shock itself.  In the case of
M16, for example, the distance between the ionization front and the
ionization shock is observed to be about 0.2 pc \citep{fhs02}, while
the shock speed is about 1.3 km s$^{-1}$.  A dense core overrun by the
ionization shock will therefore be uncovered by the ionization front
within about $1.5 \times 10^5$ years.  Maser emission requires that
the protostar be embedded within dense molecular material
\citep{eli92}.  This is both expected theoretically, and borne out by
the fact that no masers are observed around sources located in the low
density interiors of \Htwo\ regions.  So, as the ionization front
arrives, the maser-exciting lifetime of a protostar will come to an
end.

A lot must happen in the $\sim 2 \times 10^5$ years between the triggering of
star formation and the arrival of the ionization front.  In isolation,
for example, the free fall time for a 1 \Msun\ core with a temperature
of 10 K is about $5 \times 10^5$ years.  Many and perhaps most
potentially star-forming cores will be overrun before they form stars
at all.  These could account for some of the EGGs that are not seen to
contain 2 $\micron$ sources.  Other ``empty'' EGGs may contain
protostars and brown dwarfs that are simply not yet visible at
2$\micron$.  Other protostars might have just begun their
maser-producing phase when they are uncovered.  If this mechanism is
the correct explanation for our observations, it requires that the
typical protostar in M16 has its maser-producing lifetime shortened by
a factor of ten.  This number is not out of line with the fact that
only about 15\% of EGGs are known to have reached the point of forming
protostars at all.

If our interpretation is correct, it has significant implications for
the process of star formation in \Htwo\ region environments.  It means
that the objects observed by \citep{ma01} as they emerge from the
columns in M16 were still potentially maser-producing protostars prior
to exposure.  In other words, it seems likely that these objects were
still accreting mass up until the time they were overrun by the
ionization front, as originally proposed by \citep{hes96}.

A possible objection to the star formation scenario discussed here is
that the accretion luminosity associated with triggered star formation
should be seen.  In the scenario that we propose, a forming protostar
must radiate away its binding energy in the $\sim 2 \times 10^5$ years
between the triggering of collapse and the subsequent disruption of
the region by the ionization front.  This implies that a protostar
must have an accretion luminosity of $\sim$ 100 $L_{\odot}$.  We
require that M 16 Column 2, for example, contain perhaps 25
protostars.  The ratio of the submillimeter luminosity of a protostar
to its bolometric luminosity is about 0.005 \citet{and96}, so we
predict that the submillimeter luminosity from Column 2 due to
protostars should be about 13 $L_{\odot}$.  This is a factor of 30
less than the $\sim 400 L_{\odot}$ submillimeter luminosity of Column
2 observed by \citet{whi99}.  In other words, the submillimeter
luminosity due to protostars would be hard to recognize against the
background luminosity of the dust in the columns.  This is true even
if protostars radiate a much larger fraction of their luminosity at
submillimeter wavelengths.

An embedded protostar will not be visible in the near infrared, and
its accretion luminosity will fall off rapidly once it is uncovered by
the ionization front.  There is only a brief window of time, just as
the protostar emerges from the molecular cloud, when the accretion
luminosity of the protostar might be observed directly.  It is
therefore significant that \citet{tsh02} observed two such luminous
protostars (ES-1 and ES-2, which have luminosities of 200 $L_{\odot}$
and 20 $L_{\odot}$, respectively) sitting precisely at the ionization
fronts in Columns 1 and 2.  Two other luminous protostars (M16S-1 and
M16S-2) with luminosities of 20 $L_{\odot}$ are located just outside
of the ionization front in Column 3.  It is hard to interpret these
objects as anything other than protostars that were accreting up until
the time they were overrun by the ionization front.

It is interesting to speculate about the role of triggering and
subsequent disruption of star formation acting on larger scales and in
more extreme environments. \citet{sco98} conducted a detailed
comparison between the physical environments at ionization boundaries
in 30 Doradus and the physical environments seen in local regions such
as M16.  Their conclusion was that conditions in these two types of
regions are quite similar.  The same ranges of pressures,
temperatures, ionization stratification, and physical structures are
present in both 30 Dor and M16.  The difference between the two is
simply a matter of overall scale.  The much greater total flux from 30
Dor is able to impose these conditions over a much larger volume.  The
same mechanisms responsible for star formation in M 16 should then
also be at work in giant \Htwo\ regions.  High angular-resolution
observations of water masers in 30 Dor \citep{laz02} indeed show a
close association between the \Htwo\ region/molecular cloud boundaries
and the sites of active star formation.

\section{Summary and Conclusions}

We have searched six \Htwo\ regions for water maser emission
associated with low-mass star formation. We detected a total of 16
water maser systems, with only the emission from S140 IRS~1 reported
previously. The distribution of water masers within M16 and M20 is not
random. Instead, the masers are strongly concentrated in regions that
have recently been compressed by ionization shocks, and will be
overrun by ionization fronts on timescales comparable to the
maser-exciting lifetimes of protostars.  We have also found that the
fraction of protostars that excite masers is significantly lower in
\Htwo\ region environments than in regions of isolated star formation.

The original motivation for this work was in part to test a scenario
for star formation proposed by \citet{hes96} in their work on M16.
Radiative energy from massive O and B stars is deposited at the
surface of adjacent molecular clouds, where it drives a stratified
photoionized photoevaporative flow away from the interface
\citep[][and references therein]{sh00}, and drives an ionization
shock into the molecular gas.  Ongoing star formation that is
originally triggered by the ionization shock is disrupted a short time
later by the advance of the ionization front.  Stars forming as a
result of this process will pass through a well-defined series of
stages, beginning as sub-Jeans cores, and ending as exposed young
stellar objects, adrift within the ionized interior of the \Htwo\
region.

We found that the distribution of maser emission in \Htwo\ regions
allowed us to test two specific predictions of this scenario for star
formation: (1) low-mass star formation, if it is triggered by
compression in advance of ionization fronts, should be localized
near ionization fronts; and (2) low-mass stars should spend less time
enshrouded in high density material in \Htwo\ region environments than
they do in regions of isolated star formation.  The findings of the
present study are in accord with these predictions.

Much work remains before we can claim to have sorted out the complex
process of star formation in \Htwo\ region environments.  However, our
data leave little doubt that pockets of star formation are triggered
around \Htwo\ regions, and that the accretion phase of the star
formation process proceeds differently in \Htwo\ region environments
than in regions of isolated low-mass star formation.  Since most low
mass stars form in \Htwo\ region environments, understanding the
details of this process is of great importance in answering a diverse
set of questions ranging from the origins of the IMF to the
environments in which planet formation occurs.

\acknowledgments

The authors would like to acknowledge useful conversations with Steve
Desch.  This work has been supported in part by grants
HST-GO-06574.01-A and HST-GO-09091.01-A from the Space Telescope
Science Institute.  K.R.H. gratefully acknowledges support from the
Arizona NASA Space Grant. This publication makes use of data products
from the Two Micron All Sky Survey, which is a joint project of the
University of Massachusetts and the Infrared Processing and Analysis
Center/California Institute of Technology, funded by the National
Aeronautics and Space Administration and the National Science
Foundation. This research has made use of the NASA/IPAC Infrared
Science Archive, which is operated by the Jet Propulsion Laboratory,
California Institute of Technology, under contract with the National
Aeronautics and Space Administration.

\clearpage
\begin{deluxetable}{cccrc}
\tablecaption{Observations of \Htwo\ Regions \label{observed}}
\tablewidth{0pt}
\tablehead{
\colhead{Name} & \colhead{$\alpha$ (2000)} & \colhead{$\delta$ (2000)} &
\colhead{Observation Dates} & \colhead{Velocity Range (\kms)}
}
\startdata
NGC~2264 & 06 41 12.80 &   +09 28 16.0 &    2000 Apr 04, Jun 13 & $-$14 to $+$28\\
         & 06 41 19.82 &   +09 27 16.0 &    (C configuration)   &        \\
         & 06 41 05.78 &   +09 27 16.0 &                        &        \\
         & 06 41 12.80 &   +09 26 16.0 &                        &        \\
         & 06 41 19.82 &   +09 25 16.0 &                        &        \\
         & 06 41 05.78 &   +09 25 16.0 &                        &        \\
         & 06 41 12.80 &   +09 24 16.0 &                        &        \\
         &             &               &                        &        \\
NGC~6357 & 17 24 51.21 & $-$34 08 48.3 &    2001 Mar 26, May 18 & $-$5 to $+$37\\
         & 17 24 41.55 & $-$34 08 48.3 &    (B configuration)   &        \\
         & 17 24 36.71 & $-$34 10 32.2 &                        &        \\
         & 17 24 46.38 & $-$34 10 32.2 &                        &        \\
         & 17 24 56.05 & $-$34 10 32.2 &                        &        \\
         & 17 24 51.21 & $-$34 12 16.1 &                        &        \\
         & 17 24 41.55 & $-$34 12 16.1 &                        &        \\
         &             &               &                        &        \\
M16      & 18 18 43.91 & $-$13 46 46.5 &    2000 Apr 02, Jun 06 & $+$4 to $+$46\\
         & 18 18 40.93 & $-$13 48 39.0 &    (C configuration)   &        \\
         & 18 18 49.16 & $-$13 48 21.0 &                        &        \\
         & 18 18 46.08 & $-$13 50 10.5 &                        &        \\
         & 18 18 54.21 & $-$13 49 57.0 &                        &        \\
         & 18 18 50.91 & $-$13 51 48.0 &                        &        \\
         & 18 18 59.15 & $-$13 51 34.5 &                        &        \\
         &             &               &                        &        \\
         & 18 19 08.00 & $-$13 44 56.2 &    2001 Mar 26, May 18 & $+$4 to $+$46\\
         & 18 18 53.59 & $-$13 44 44.2 &    (B configuration)   &        \\
         & 18 18 45.35 & $-$13 44 44.2 &                        &        \\
         & 18 18 49.47 & $-$13 46 28.1 &                        &        \\
         & 18 18 51.53 & $-$13 49 32.2 &                        &        \\
         & 18 18 56.48 & $-$13 51 08.0 &                        &        \\
         & 18 19 14.60 & $-$13 50 19.4 &                        &        \\
         & 18 19 08.02 & $-$13 51 31.7 &                        &        \\
         & 18 19 09.03 & $-$13 53 30.8 &                        &        \\
         & 18 19 01.44 & $-$13 52 43.9 &                        &        \\
         & 18 19 02.45 & $-$13 54 43.0 &                        &        \\
         & 18 18 54.86 & $-$13 53 56.1 &                        &        \\
         & 18 18 55.87 & $-$13 55 55.2 &                        &        \\
         &             &               &                        &        \\
         & 18 18 58.83 & $-$13 52 47.1 &    2002 Mar 14         & $-$15 to $+$65\\
         & 18 19 03.72 & $-$13 52 03.7 &    (A configuration)   &        \\
         & 18 18 48.90 & $-$13 49 50.0 &                        &        \\
         & 18 18 50.42 & $-$13 48 51.8 &                        &        \\
         & 18 18 46.20 & $-$13 44 27.0 &                        &        \\
         & 18 18 48.71 & $-$13 45 20.9 &                        &        \\
         & 18 19 07.10 & $-$13 45 25.0 &                        &        \\
         & 18 19 24.95 & $-$13 45 40.3 &                        &        \\
         &             &               &                        &        \\
M20      & 18 02 32.24 & $-$23 01 51.0 &    2000 Apr 02, Jun 06 & $-$2 to $+$40\\
         & 18 02 23.55 & $-$23 01 51.0 &    (C configuration)   &        \\
         & 18 02 36.59 & $-$23 03 35.0 &                        &        \\
         & 18 02 27.90 & $-$23 03 35.0 &                        &        \\
         & 18 02 19.20 & $-$23 03 35.0 &                        &        \\
         & 18 02 32.24 & $-$23 05 19.0 &                        &        \\
         & 18 02 23.55 & $-$23 05 19.0 &                        &        \\
         &             &               &                        &        \\
         & 18 02 33.20 & $-$22 58 05.4 &    2001 Mar 26, May 18 & $-$2 to $+$40\\
         & 18 02 24.51 & $-$22 58 05.4 &    (B configuration)   &        \\
         & 18 02 28.86 & $-$23 03 43.1 &                        &        \\
         & 18 02 11.47 & $-$23 03 43.1 &                        &        \\
         & 18 02 07.12 & $-$23 05 27.1 &                        &        \\
         & 18 02 15.81 & $-$23 05 27.1 &                        &        \\
         & 18 02 24.51 & $-$23 05 27.1 &                        &        \\
         & 18 02 33.20 & $-$23 05 27.1 &                        &        \\
         & 18 02 48.42 & $-$23 05 06.3 &                        &        \\
         & 18 02 37.55 & $-$23 07 11.0 &                        &        \\
         & 18 02 28.86 & $-$23 07 11.0 &                        &        \\
         & 18 02 20.16 & $-$23 07 11.0 &                        &        \\
         &             &               &                        &        \\
         & 18 02 06.00 & $-$23 05 27.0 &    2002 Mar 14         & $-$21 to $+$59\\
         & 18 02 30.92 & $-$23 05 49.0 &    (A configuration)   &        \\
         & 18 02 28.47 & $-$23 03 49.0 &                        &        \\
         &             &               &                        &        \\
S125     & 21 53 49.27 &   +47 17 59.9 &    2001 Mar 26, May 18 & $-$13 to $+$29\\
         & 21 53 49.27 &   +47 15 59.9 &    (B configuration)   &        \\
         & 21 53 49.27 &   +47 13 59.9 &                        &        \\
         & 21 53 39.06 &   +47 12 59.9 &                        &        \\
         & 21 53 39.06 &   +47 14 59.9 &                        &        \\
         & 21 53 39.06 &   +47 16 59.9 &                        &        \\
         & 21 53 39.06 &   +47 18 59.9 &                        &        \\
         & 21 53 28.85 &   +47 19 59.9 &                        &        \\
         & 21 53 28.85 &   +47 17 59.9 &                        &        \\
         & 21 53 28.85 &   +47 15 59.9 &                        &        \\
         & 21 53 28.85 &   +47 13 59.9 &                        &        \\
         & 21 53 28.85 &   +47 11 59.9 &                        &        \\
         & 21 53 18.64 &   +47 12 59.9 &                        &        \\
         & 21 53 18.64 &   +47 14 59.9 &                        &        \\
         & 21 53 18.64 &   +47 16 59.9 &                        &        \\
         & 21 53 18.64 &   +47 18 59.9 &                        &        \\
         & 21 53 08.43 &   +47 17 59.9 &                        &        \\
         & 21 53 08.43 &   +47 15 59.9 &                        &        \\
         & 21 53 08.43 &   +47 13 59.9 &                        &        \\
         &             &               &                        &        \\
S140     & 22 18 50.88 &   +63 17 18.0 &           2000 June 06 & $-$33 to $+$9\\
         & 22 19 01.56 &   +63 18 55.5 &    (C configuration)   &        \\
         & 22 19 08.90 &   +63 17 04.5 &                        &        \\
         & 22 19 19.35 &   +63 18 45.0 &                        &        \\
         & 22 19 26.91 &   +63 16 54.0 &                        &        \\
         & 22 19 37.14 &   +63 18 33.0 &                        &        \\
         & 22 19 44.71 &   +63 16 40.5 &                        &
\enddata
\end{deluxetable}

\begin{deluxetable}{lllrr}
\tablecaption{VLA Calibrators \label{calib}}
\tablewidth{0pt}
\tablehead{
\colhead{Name} & \colhead{Source} & \colhead{Obs. Date} &
\colhead{Configuration} & \colhead{Flux Density (Jy)}
}
\startdata
0137$+$331 & primary cal.  & 2000 Apr 02 & C & 1.19  \\
           & primary cal.  & 2000 Apr 04 & C & 1.19  \\
           & primary cal.  & 2000 Jun 06 & C & 1.19  \\
           & primary cal.  & 2000 Jun 13 & C & 1.19  \\
           & primary cal.  & 2001 Mar 26 & B & 1.32  \\
           & primary cal.  & 2001 May 18 & B & 1.31  \\
           &               &             &   &       \\
1331$+$305 & primary cal.  & 2002 Mar 14 & A & 2.54  \\
           &               &             &   &       \\
0700$+$171 & NGC~2264      & 2000 Apr 04 & C & 0.95  \\
           & NGC~2264      & 2000 Jun 13 & C & 0.84  \\
           &               &             &   &       \\
1733$-$130 & M16           & 2000 Apr 02 & C & 3.61  \\
1733$-$130 & M16           & 2000 Jun 06 & C & 4.15  \\
1733$-$130 & M16, NGC~6357 & 2001 Mar 26 & B & 8.37  \\
1733$-$130 & M16, NGC~6357 & 2001 May 18 & B & 6.99  \\
1733$-$130 & M16           & 2002 Mar 14 & A & 7.37  \\
           &               &             &   &       \\
1820$-$254 & M20           & 2002 Mar 14 & A & 0.78  \\
           &               &             &   &       \\
1911$-$201 & M20           & 2000 Apr 02 & C & 3.89  \\
1911$-$201 & M20           & 2000 Jun 06 & C & 3.46  \\
1911$-$201 & M20           & 2001 Mar 26 & B & 3.95  \\
1911$-$201 & M20           & 2001 May 18 & B & 3.33  \\
           &               &             &   &       \\
2022$+$616 & S140          & 2000 Jun 06 & C & 2.23  \\
           &               &             &   &       \\
2202$+$422 & S125          & 2001 Mar 26 & B & 3.57  \\
2202$+$422 & S125          & 2001 May 18 & B & 3.01
\enddata
\end{deluxetable}

\clearpage

\setcounter{page}{29}

\begin{deluxetable}{ccclrr}
\tablecaption{Water Maser Properties \label{detected}}
\rotate
\tablewidth{0pt}
\tablehead{
\colhead{Name} & \colhead{$\alpha$ (2000)} & \colhead{$\delta$ (2000)} &
\colhead{Date} & \colhead{Peak Flux Density(Jy)} & \colhead{Peak Velocity (\kms)}
}
\startdata
M16 Col. 2 & 18 18 48.8950 & $-$13 49 50.100 & 2000 Apr 02 & 14.9  &  +21.4 \\
           & 18 18 48.8950 & $-$13 49 50.100 & 2000 Jun 06 & 20.7  &  +22.0 \\
           & 18 18 48.8871 & $-$13 49 49.950 & 2001 Mar 26 &  1.17, 0.38, 0.50
                                                           &  +14.8, +18.7, +22.7 \\
           & 18 18 48.9186 & $-$13 49 50.429 & 2001 May 18 &  0.84 &  +23.4 \\
           & 18 18 48.8918 & $-$13 49 50.075 & 2002 Mar 14 &  0.13 &  +21.8 \\
           &               &                 &             &       &        \\
M16 Col. 4 & 18 18 58.8341 & $-$13 52 47.300 & 2000 Apr 02 &  2.71 &  +25.7 \\
           & 18 18 58.8341 & $-$13 52 47.500 & 2000 Jun 06 &  0.68, 1.06 
                                                           &  +24.7, +28.0 \\
           & 18 18 58.8193 & $-$13 52 47.123 & 2001 Mar 26 & 10.6, 0.45  
                                                           &  +26.0, +28.3 \\
           & 18 18 58.8564 & $-$13 52 47.843 & 2001 May 18 &  1.52, 0.41 
                                                           &  +25.7, +26.3 \\
           & 18 18 58.8228 & $-$13 52 47.190 & 2002 Mar 14 &  0.49 &  +23.4 \\

M16 Col. 5 &               &                 &             &       &        \\
         A & 18 19 07.0908 & $-$13 45 24.590 & 2001 Mar 26 &  6.74, 2.97 
                                                           &  +16.4, +21.1 \\
         A & 18 19 07.1567 & $-$13 45 24.770 & 2001 May 18 &  2.74, 1.48 
                                                           &  +16.4, +20.7 \\
         A & 18 19 07.1233 & $-$13 45 24.600 & 2002 Mar 14 &  3.72, 5.31 
                                                           &  +16.2, +20.8 \\
         B & 18 19 07.0008 & $-$13 45 21.375 & 2002 Mar 14 &  0.19 &  +32.2 \\
         C & 18 19 07.3375 & $-$13 45 23.123 & 2002 Mar 14 &  1.49 &  +26.9 \\

M16 N Bay  &               &                 &             &       &        \\
         A & 18 18 46.4247 & $-$13 44 25.730 & 2001 Mar 26 &  1.18, 6.31 
                                                           &  +15.7, +24.7 \\
         A & 18 18 46.4247 & $-$13 44 25.550 & 2001 May 18 &  1.79, 2.93 
                                                           &  +15.8, +24.7 \\
         B & 18 18 46.0747 & $-$13 44 30.530 & 2001 May 18 &  1.77 &  +14.8 \\
         B & 18 18 46.0943 & $-$13 44 30.835 & 2002 Mar 14 &  0.24 &   +6.3 \\
         C & 18 18 46.7130 & $-$13 44 14.165 & 2002 Mar 14 &  0.10, 0.14, 0.12 
                                                           & +7.0, +16.9, +26.2 \\
           &               &                 &             &       &        \\
M20 SE     & 18 02 30.9210 & $-$23 05 49.200 & 2000 Apr 02 &  0.59 &  +24.9 \\
           & 18 02 30.9210 & $-$23 05 49.200 & 2000 Jun 06 &  0.24 &  +25.3 \\

M20 TC3    &               &                 &             &       &        \\
         A & 18 02 05.7924 & $-$23 05 25.022 & 2001 Mar 26 &  0.63, 0.32, 0.28 
                                                           & +5.5, +20.3, +25.3 \\
         A & 18 02 05.7924 & $-$23 05 24.902 & 2001 May 18 &  0.63, 0.43, 0.24 
                                                           & +5.2, +20.0, +25.6 \\ 
         A & 18 02 05.7880 & $-$23 05 24.990 & 2002 Mar 14 &  0.42, 2.81, 0.77, 
                                                           &$-$12.2, +16.1, +20.2, \\
           &               &                 &             &  6.43, 0.22 
                                                           & +25.8 +33.7 \\
         B & 18 02 05.5880 & $-$23 05 29.162 & 2001 Mar 26 &  0.38 &   +2.5 \\
         B & 18 02 05.5804 & $-$23 05 29.205 & 2002 Mar 14 &  2.44, 22.7, 0.42 
                                                           &$-$14.8, $-$11.8, $-$3.6 \\
         C & 18 02 06.2446 & $-$23 05 23.402 & 2001 May 18 &  0.35 &  +24.9 \\
           &               &                 &             &       &        \\
NGC~2264   & 06 41 12.2593 &   +09 29 12.000 & 2000 Apr 04 &  0.57 &   +2.6 \\
           & 06 41 12.2593 &   +09 29 11.800 & 2000 Jun 13 &  0.43 &   +2.6 \\

S140       &               &                 &             &       &        \\
         A & 22 19 18.2219 &   +63 18 46.800 & 2000 Jun 06 &  0.59, 2.37 
                                                           &$-$15.3, +3.5 \\
         B & 22 19 16.0846 &   +63 18 38.798 & 2000 Jun 06 &  1.46 &$-$17.9 \\
         C & 22 19 21.1314 &   +63 18 54.599 & 2000 Jun 06 &  0.96 & $-$5.7
\enddata
\end{deluxetable}

\clearpage

\setcounter{page}{31}

\clearpage

\clearpage

\clearpage

\clearpage

\clearpage

\clearpage

\clearpage

\end{document}